\documentclass[a4paper,12pt]{article}
\usepackage{hyperref}
\renewcommand{\theequation}{\arabic{section}.\arabic{equation}}
\renewcommand{\thefootnote}{\fnsymbol{footnote}}
\begin{document} 
\begin{titlepage}
\null \hfill hep-th/0510088\\
\null \hfill Preprint TU-756  \\
\null \hfill October 2005 \\[4ex]

\begin{center}
{\LARGE Instantons in ${\cal N}=1/2$ Super Yang-Mills Theory \\
via Deformed Super ADHM Construction
}\\[3em] 
%%%%%%%%%% Here is the title of this Note
%\today

{\large %%%%%%%%%%%%%%%%%% here authors name %%%%%%%%%%%%
Takeo Araki\footnote{araki@tuhep.phys.tohoku.ac.jp}, \ 
Tatsuhiko Takashima\footnote{takashim@tuhep.phys.tohoku.ac.jp}\ \ and \ %\\
Satoshi Watamura\footnote{watamura@tuhep.phys.tohoku.ac.jp}\\ [2ex]
}
%%%%%%%%%%%%%%%%%%%%%%%%%%%%%% address of the institute %%%%%%%%
Department of Physics \\
Graduate School of Science \\
Tohoku University \\
Aoba-ku, Sendai 980-8578, Japan \\ [2ex]

\vspace*{1cm} 
%%%%%%%%%%%%%%%%%%%%%%%%%%%%% This is abstruct %%%%%%%%%%%%%%%

\begin{abstract}
We study an extension of the ADHM construction 
to give deformed anti-self-dual (ASD) instantons 
in ${\cal N}=1/2$ super Yang-Mills theory 
with ${\rm U}(n)$ gauge group. 
First we extend the exterior algebra on superspace 
to non(anti)commutative superspace 
and show that the ${\cal N}=1/2$ super Yang-Mills theory 
can be reformulated in a geometrical way. 
By using this exterior algebra, we formulate 
a non(anti)commutative version of the super ADHM construction 
and show that the curvature two-form superfields 
obtained by our construction do satisfy the deformed ASD equations
and thus we establish the deformed super ADHM construction. 
We also show that the known deformed ${\rm U}(2)$ one instanton solution 
is obtained by this construction. 
\end{abstract}

\end{center}

\end{titlepage}

%\tableofcontents
\renewcommand{\thefootnote}{\arabic{footnote}}
\setcounter{footnote}{0} 
\section{Introduction}\setcounter{equation}{0}

In supersymmetric Yang-Mills theories, 
there are zero modes of adjoint fermions 
in the instanton background. 
Their existence naturally introduces 
the superpartner of the bosonic moduli 
called Grassmann collective coordinates (or fermionic moduli). 
The fermion zero modes together with the bosonic configurations 
are called super instantons. 
For reviews, see refs.\ \cite{DoHoKhMa,KhMaSl} for example.  

It is well known that the instanton configurations of the gauge field 
can be obtained by the ADHM construction \cite{AtHiDrMa}. 
To give the super instanton solutions, 
superfield extensions of the ADHM construction were proposed in \cite{Sem,Vo} 
(see also \cite{McAr1}-\cite{DeOg2}), 
in which the fermionic moduli belong to the same superfield 
containing the bosonic moduli. 
In the previous paper \cite{ArTaWa}, 
we formulated the ${\cal N}=1$ super ADHM construction 
with the use of the $\ddagger$-conjugation, 
and found a condition to ensure the Wess-Zumino (WZ) gauge of 
the gauge potential superfield (as well as the field strength superfields) 
obtained by this ADHM construction. 
The investigation of the WZ gauge is necessary 
if one would like to compare the results in the superfield formalism 
with those obtained by the component formalism. 
Especially it is indispensable for our present formulation of the
deformed super ADHM construction in the following.
 
One of the motivations of our previous paper is 
emergence of supersymmetric gauge theory defined 
on a kind of deformed superspace, 
called non(anti)commutative superspace, 
in superstring theory 
as a low energy effective theory on D-branes 
with constant graviphoton field strength 
\cite{OoVa}-\cite{BeSe} 
(see \cite{NAC} for earlier works on deformed superspace). 
In non(anti)commutative space, 
anticommutators of Grassmann coordinates become non-vanishing. 
Such a deformation of (Euclidean) four dimensional 
${\cal N}=1$ super Yang-Mills theory has been formulated by Seiberg \cite{Se}, 
which is sometimes called ${\cal N}=1/2$ super Yang-Mills theory. 
Subsequently non(anti)commutative gauge theories have been studied extensively 
in both perturbative and non-perturbative aspects \cite{pert}-\cite{ItNa}. 
 
It was argued by Imaanpur \cite{Im} 
that the anti-self-dual (ASD) instanton equations should be modified 
in the ${\cal N}=1/2$ super Yang-Mills theory 
with self-dual (SD) non(anti)-commutativity. 
Solutions to those equations (deformed ASD instantons) have been studied 
by many authors \cite{Im}-\cite{GiRiRoTr} (see also \cite{ItNa}). 
In the case of ${\rm U}(2)$ gauge group, 
the exact one-instanton solution have been explicitly constructed 
in \cite{Im,GrRiRo}  
by perturbation with respect to the non-anticommutativity parameter. 
${\rm U}(n)$ ($n\ge2$) one-instanton solutions are obtained in \cite{BrFeLuRe} 
in a similar way. 
In ref.\ \cite{BiFrPeLe}, the authors have studied string amplitudes 
in the presence of D($-1$)-D$3$ branes with the background R-R field strength 
and derived constraint equations for the string modes ending 
on D($-1$)-branes, 
which are nothing but the ADHM constraints for the deformed ASD instantons. 
On the other hand, it is far from obvious how to obtain these constraints 
in the purely field theoretic context 
and how the deformed ASD connections are given exactly
in terms of the ADHM moduli parameters. 
Clearly we need an appropriately extended ADHM construction 
to answer these questions. 
Then it is natural to expect that useful is 
the superfield extension of the ADHM construction, 
because the field theories on non(anti)commutative superspace can be realized by deforming 
the multiplication of superfields. 
 
In this paper, we extend the ADHM construction 
to the one that can give exact solutions 
to the deformed ASD equations 
in ${\cal N}=1/2$ super Yang-Mills theory with ${\rm U}(n)$ gauge group. 
This is accomplished by deforming the ${\cal N}=1$ super ADHM construction 
which we have studied in the previous paper. 
Our formulation provides a way to obtain 
other possible solutions beyond the one instanton configurations. 
 
This paper is organized as follows. 
In section \ref{sect:N=1/2SYM}, 
we review ${\cal N}=1/2$ super Yang-Mills theory and 
the deformed ASD equations. 
In section \ref{sect:ExteriorAlgebra}, 
we define a deformed exterior algebra on the non(anti)commutative superspace 
and show that the ${\cal N}=1/2$ super Yang-Mills theory 
can be reproduced in a geometrical way, based on this
 deformed exterior algebra. 
In section \ref{sect:DSADHM}, 
we describe a non(anti)commutative version of 
the ${\cal N}=1$ super ADHM construction 
after briefly reviewing the undeformed super ADHM construction. 
We show that the curvature two-form superfields obtained 
by our construction do satisfy the deformed ASD equations. 
Section \ref{sect:Conclusion} is devoted to conclusions and discussion. 
In appendix \ref{app:Notation}, 
we describe a few of our notation and conventions, 
although we follow our previous paper \cite{ArTaWa} 
\footnote{ 
We should notice, however, that 
there is a change in the notation from our previous paper. 
The ``anti-holomorphic'' quantities with respect to the $\ddagger$-conjugation 
are indicated by ``$\widetilde{\ ^{\ }}$'' in ref.\ \cite{ArTaWa}, 
while they are indicated simply by ``${}^\ddagger$'' in this paper. 
For example, $\widetilde{\hat{\Delta}}{}_\alpha$ in \cite{ArTaWa} 
is denoted as $\hat{\Delta}^\ddagger{}_\alpha$. 
}. 
In appendix \ref{app:Inverse}, we give the ``inverse'' 
of a chiral superfield with respect to the star product, 
which is needed in formulating the deformed super ADHM construction. 
In appendix \ref{app:Determination}, we give a detailed derivation of 
the normalized zero mode superfield of the zero-dimensional Dirac operator. 
In appendix \ref{app:U(2)OneInstanton}, 
we obtain the known ${\rm U}(2)$ one instanton solution 
by the deformed super ADHM construction.

\section{Non(anti)commutative deformation of ${\cal N}=1$ super Yang-Mills}
\setcounter{equation}{0}
\label{sect:N=1/2SYM}

We will briefly describe 
the non(anti)commutative deformation of ${\cal N}=1$ superspace and 
${\cal N}=1/2$ super Yang-Mills theory 
formulated in \cite{Se}.

The non(anti)commutative deformation of ${\cal N}=1$ superspace is given by 
introducing non(anti)commutativity 
of the product of ${\cal N}=1$ superfields. 
This deformation is realized by the following star product:  
\begin{equation}
 f*g=f \exp(P) g ,\quad
P=-\frac12
\overleftarrow{Q_{\alpha}}
C^{\alpha\beta}\overrightarrow{Q_{\beta}}, 
\label{eq:StarProduct1}
\end{equation}
where $f$ and $g$ are ${\cal N}=1$ superfields and 
$Q_\alpha$ is the (chiral) supersymmetry generator. 
$C^{\alpha\beta}$ is the non-anticommutativity parameter 
and is symmetric: $C^{\alpha\beta}=C^{\beta\alpha}$. 
The above star product gives the following relations 
among the chiral coordinates 
$(y^\mu, \theta^\alpha, \bar{\theta}^{\dot{\alpha}})$: 
\begin{equation}
 \{ \theta^{\alpha}, \theta^{\beta}\}_{*}
= 
	C^{\alpha\beta}
	, \quad 
 {[ y^{\mu}, \cdot \, ]}_{*}
= 
	0 
	, \quad 
 {[} \bar{\theta}^{\dot{\alpha}} , \cdot \, \}_{*}
= 
	0. 
\end{equation}
In terms of the coordinates 
$(x^\mu, \theta^\alpha, \bar{\theta}^{\dot{\alpha}})$, 
these relations are  
\begin{equation}
 \{ \theta^{\alpha}, \theta^{\beta}\}_{*}
= 
	C^{\alpha\beta}
	, \quad 
{[ x^{\mu}, x^{\nu} ]}_{*} 
=
	C^{\mu\nu} \bar{\theta} \bar{\theta}
	, \quad 
\mbox{[} x^{\mu},\theta^{\alpha} \mbox{]}_{*}
= 
	i C^{\alpha\beta}
		(\sigma^{\mu}\bar{\theta})_{\beta}
	, \quad 
{[} \bar{\theta}^{\dot{\alpha}} , \cdot \, \}_{*}
= 
	0 
	, 
\end{equation}
where
\begin{equation}
C^{\mu\nu}\equiv C^{\alpha\beta}
(\sigma^{\mu\nu})_{\alpha} {}^{\gamma} \varepsilon_{\beta\gamma}
. 
\end{equation}
Since $P$ commutes with the supercovariant spinor derivatives 
$D_\alpha$ and $\bar{D}_{\dot{\alpha}}$, 
the chirality notion of superfields is preserved. 
For example, given two chiral superfields $\Phi_i(y,\theta)$ 
($i=1,2$), the product $\Phi_1 * \Phi_2$ becomes another chiral superfields. 
Turning on such a deformation, 
the original action formulated in the ${\cal N}=1$ superfield formalism 
is deformed by the star product. 
Since $P$ also commutes with $Q_\alpha$, 
the deformed action preserves in general 
the chiral half of the supersymmetry transformation 
generated by $Q_\alpha$. 
The deformed ${\cal N}=1$ super Yang-Mills theory 
has ${\cal N}=1/2$ supersymmetry, so that they are called ${\cal N}=1/2$ 
super Yang-Mills theory.

The $\ddagger$-conjugation \cite{ArTaWa} of $C^{\alpha\beta}$ 
is deduced from the $\ddagger$-conjugation of $\theta^\alpha \theta^\beta$ 
and found as 
\begin{equation} 
( C^{\alpha\beta} )^\ddagger 
= 
	- C_{\beta\alpha} 
	, 
\end{equation} 
where 
$C_{\alpha\beta} 
\equiv 
	\varepsilon_{\alpha\gamma} \varepsilon_{\beta\delta} C^{\gamma\delta}
$. 
Let $A$ and $B$ are superfields. 
We define the $\ddagger$-conjugation of 
$Q_\alpha (A)$ as 
\begin{equation} 
( Q_\alpha (A) )^\ddagger 
\equiv 
	(-)^{|A|} (A)^\ddagger \overleftarrow{Q^\alpha} 
	,
\end{equation}
then we have 
\begin{equation} 
( A * B )^\ddagger 
= 
	(-)^{|A||B|} (B)^\ddagger * (A)^\ddagger 
. 
\end{equation}

The action of ${\cal N}=1/2$ supersymmetric Yang-Mills theory 
is given by 
\begin{equation}
 S= \frac{1}{16 N g^2}\int d^4x\left(
	\int d^2\theta {\rm tr} W^\alpha*W_\alpha
	+ \int d^2\bar{\theta}
                       {\rm tr} \bar{W}_{\dot{\alpha}}*\bar{W}^{\dot{\alpha}}
                           \right)
\label{eq:C:Action} 
\end{equation} 
where 
\begin{equation}
W_\alpha
=
	-\frac{1}{4}\bar{D}_{\dot{\alpha}}\bar{D}^{\dot{\alpha}} 
		\left(  e_*^{-V} * D_\alpha e_*^V \right)
	, \quad 
\bar{W}_{\dot{\alpha}}
=
	\frac{1}{4}D^\alpha D_\alpha 
		\left(e_*^V * \bar{D}_{\dot{\alpha}}e_*^{-V}\right)
	, 
	\label{eq.Wdeform}
\end{equation}
and $e_*^V \equiv \sum_n \frac{1}{n!} \overbrace{V*\cdots *V}^n$. 
Here $V=V^a T^a$ with $V^a$ the vector superfields and 
$T^a$ the hermitian generators 
which are normalized as ${\rm tr}[ T^a T^b ] = N \delta^{ab}$. 
We may redefine the component fields of $V$ in the WZ gauge 
such that the component gauge transformation becomes canonical 
(the same as the undeformed case). 
In \cite{Se}, such a field redefinition is found to be 
\begin{eqnarray}
 V_{\mathrm{WZ}}(y,\theta,\bar{\theta})
&=& 
	- \theta\sigma^{\mu}\bar{\theta} v_{\mu}(y)
	+ i \theta\theta\bar{\theta}\bar{\lambda}(y)
	- i \bar{\theta}\bar{\theta}\theta^{\alpha}
		\Bigl(
		\lambda_{\alpha}
		+ {1\over4}\varepsilon_{\alpha\beta}C^{\beta\gamma}
			\sigma^{\mu}_{\gamma\dot{\gamma}}
			\left\{ \bar{\lambda}^{\dot{\gamma}},v_{\mu} \right\}
		\Bigr) (y)
	\nonumber\\
&&{} 
	+ {1\over2} \theta\theta\bar{\theta}\bar{\theta}
		\left( 
		D
		- i \partial_{\mu} v^{\mu} 
		\right) (y) 
\label{eq.c.Vwz}
\end{eqnarray}
and then $W$ and $\bar{W}$ become 
\begin{eqnarray}
W_\alpha
&=& 
	- i \lambda_\alpha (y) 
	+ \left[ 
		\delta_\alpha^\gamma D 
		- i ( \sigma^{\mu\nu} )_{\alpha}{}^{\gamma} 
			\left( 
			v_{\mu\nu} 
			+ \frac i2 C_{\mu\nu} \bar{\lambda}\bar{\lambda} 
			\right) 
		\right] (y) \theta_\gamma 
	+ \theta\theta ( \sigma^\mu {\cal D}_\mu \bar{\lambda} )_\alpha (y) 
	, 
	\label{eq:C:WComp}
	\\ 
\bar{W}_{\dot{\alpha}}
&=& 
	i \bar{\lambda}_{\dot{\alpha}} (\bar{y}) 
	+ \bar{\theta}_{\dot{\gamma}} \Bigl\{ 
		\delta^{\dot{\gamma}}_{\dot{\alpha}} D 
		- i ( \bar{\sigma}^{\mu\nu} \varepsilon )^{\dot{\gamma}}{}_{\dot{\alpha}} 
			v_{\mu\nu} 
		\Bigr\} (\bar{y}) 
	+ \bar{\theta}\bar{\theta} 
		\left[ 
		( {\cal D}_\mu \lambda \sigma^\mu )_{\dot{\alpha}} 
		\right. \nonumber\\ 
&&\left. {} 
		- \frac12 C^{\mu\nu} \{ v_{\mu\nu} , \bar{\lambda}_{\dot{\alpha}} \} 
		- C^{\mu\nu} \{ v_\nu, {\cal D}_{\mu} \bar{\lambda}_{\dot{\alpha}} 
			- \frac i4 [ v_\mu , \bar{\lambda}_{\dot{\alpha}} ] \} 
		- {i \over 16} |C|^2 
			\{ \bar{\lambda}\bar{\lambda} , \bar{\lambda}_{\dot{\alpha}} \}
		\right] (\bar{y}) 
, 
\label{eq:C:WBarComp}
\end{eqnarray}
where $|C|^2 \equiv C^{\mu\nu} C_{\mu\nu}$. 
The component action is 
\begin{eqnarray} 
S
= 
	{1\over 4 N g^2} {\rm tr} \int d^4x 
	\left[ 
	- \frac{1}{4} v^{\mu \nu} v_{\mu \nu} 
	- i \bar{\lambda} \bar{\sigma}^{\mu} {\cal D}_{\mu} \lambda 
	+ \frac12 D^2 
	- \frac i2 C^{\mu\nu} v_{\mu\nu}\bar{\lambda}\bar{\lambda}
	+ \frac18 |C|^2 (\bar{\lambda}\bar{\lambda})^2
	\right] 
. 
\end{eqnarray}

{}From the component action, we can see that 
the equations for SD instantons 
are unchanged compared to the undeformed case: 
\begin{equation}
 v^{\mathrm{ASD}}_{\mu\nu}=0,\ \
 \bar{\lambda}=0,\ \
 {\cal D}_\mu \bar{\sigma}^\mu \lambda=0,\ \
 D=0
. 
\label{eq:SuperInstanton} 
\end{equation}
Therefore, the SD instanton solutions 
are not affected by the deformation.

On the other hand, 
the equations for ASD instantons should be modified. 
The action can be rewritten as \cite{Im} 
\begin{equation} 
S 
= 
	{1\over 4 N g^2} {\rm tr} \int d^4x 
	\left[ 
	- \frac12 \Bigl( v^{\rm SD}_{\mu \nu} 
		+ \frac i2 C_{\mu\nu}\bar{\lambda}\bar{\lambda} 
		\Bigr)^2 
	- i \bar{\lambda} \bar{\sigma}^{\mu} {\cal D}_{\mu} \lambda 
	+ \frac12 D^2 
	+ \frac14 v^{\mu\nu} \tilde{v}_{\mu\nu} 
	\right] 
, 
\end{equation} 
where 
$\tilde{v}^{\mu\nu} \equiv \frac 12 \varepsilon^{\mu\nu\rho\sigma} v_{\rho\sigma}$. 
{}From this expression, we can see that configurations 
which satisfies the equations of motion 
and is connected to the ASD instantons when turning off the deformation 
are the solutions to 
the following deformed ASD instanton equations \cite{Im}: 
\begin{equation}
 v^{\mathrm{SD}}_{\mu\nu}+\frac{i}{2}C_{\mu\nu}\bar{\lambda}\bar{\lambda}=0,\ \
 \lambda=0,\ \
 {\cal D}_\mu \sigma^\mu \bar{\lambda}=0,\ \
 D=0
. 
\label{eq.c.sinsteq}
\end{equation} 
In principle, 
these equations 
can be solved perturbatively in terms of the deformation parameter $C$.   
At the zeroth order, 
the solutions may be given by the ordinary ADHM construction.
There are right handed fermion zero modes of 
the Dirac operator in those ASD backgrounds.
At the next order, 
the field strength receives the $O(C^1)$ correction from 
these fermion zero modes through the fermion bilinear term in the equation. 
This then cause the $O(C^1)$ correction to the fermion zero modes, 
which gives the $O(C^2)$ correction to the field strength 
again through the fermion bilinear. 
In a similar way, we can obtain the higher order corrections. 
In the following sections, we would like to extend the super ADHM construction 
to the one that can give exact solutions 
to the deformed ASD instanton equations (\ref{eq.c.sinsteq}) 
without using such a perturbative analysis with respect to $C$.

\section{Differential forms in the deformed superspace}\setcounter{equation}{0}
\label{sect:ExteriorAlgebra} 

We will take a geometrical approach to formulate the deformed super ADHM construction 
by generalizing the exterior algebra: 
we extend the star product between superfields 
to the one including differential forms in superspace. 
This is accomplished 
by considering the supercharges $Q_\alpha$ in (\ref{eq:StarProduct1})  
as generators of supertranslation as their original definition. 
We will see that the deformed exterior algebra 
consistently leads to 
the ${\cal N}=1/2$ super Yang-Mills theory introduced in the previous section.

\subsection{Deformation of the exterior algebra} 
As we stated in the beginning of this section, 
the principle of our construction of the deformed exterior algebra 
is that the operators $Q_\alpha$ appearing in the star product 
are identified with the generators of supertranslation. 
Thus, the star product of differential forms 
is defined according to the representations of supersymmetry 
they belong to. 

Since the one-form bases $e^A$ are supertranslation invariant, 
we define the action of $Q_\alpha$ on $e^A$ as 
\begin{equation} 
Q_\alpha ( e^A ) = 0 
. 
\end{equation} 
Then for a 1-form $\omega=e^A \omega_A$, 
it holds that 
\begin{equation} 
Q_\alpha ( \omega ) 
= 
	(-)^{|A|} e^A Q_\alpha ( \omega_A ) 
. 
\end{equation}

Using this action of $Q_\alpha$, 
we define the deformed wedge product of 1-forms 
$\omega$ and $\omega'$ as 
\begin{equation} 
\omega \ {\buildrel * \over \wedge } \ \omega' 
\equiv 
	\omega \wedge 
		\exp\left( -\frac12 \overleftarrow{Q_\alpha} 
			C^{\alpha\beta} \overrightarrow{Q_\beta} \right) 
		\omega' 
		, 
\label{eq:StarProduct2} 
\end{equation} 
where $\overleftarrow{Q}$ ($\overrightarrow{Q}$) acts on $\omega$ ($\omega'$) 
from the right (left) 
and the normal wedge product is taken for the resulting (transformed) 1-forms. 
Note that $\omega \overleftarrow{Q_\alpha} = (-)^{|\omega|} Q_\alpha (\omega)$. 
As a result, the product of supertranslation invariant 1-forms $e^A$ 
is the same as the ordinary wedge product: 
\begin{equation} 
e^{A_1} \ {\buildrel * \over \wedge } \ e^{A_2} \ {\buildrel * \over \wedge } \ 
	\cdots \ {\buildrel * \over \wedge } \ e^{A_p} 
= 
	e^{A_1} \wedge e^{A_2} \wedge \cdots \wedge e^{A_p} 
	. 
\end{equation} 
Hereafter we will suppress the wedge symbols.

We also define the star product between a differential form $\omega$ 
and a superfield $f$ as 
\begin{equation} 
\omega * f 
\equiv 
	\omega \exp\left( -\frac12 \overleftarrow{Q_\alpha} 
			C^{\alpha\beta} \overrightarrow{Q_\beta} \right) f 
	, \quad 
f * \omega
\equiv 
	f \exp\left( -\frac12 \overleftarrow{Q_\alpha} 
			C^{\alpha\beta} \overrightarrow{Q_\beta} \right) \omega 
	. 
\label{eq:StarProduct3} 
\end{equation} 
Then it holds that 
\begin{equation} 
[ e^A , f \}_* = 0  
\end{equation} 
where $f$ is an arbitrary superfield.

With the use of the basis $e^A$, 
a $p$-form $\omega_p$ is expanded as 
\begin{equation} 
\omega_p=e^{A_1} \cdots e^{A_p} \omega_p{}_{A_p\dots A_1}
, \label{eq:C:pForm}
\end{equation} 
where the coefficients $\omega_p{}_{A_p\dots A_1}$ are general superfields. 
In this basis, the product of the $p$- and $q$-form 
is simply given by the star product of the coefficients: 
\begin{equation} 
\omega_p * \omega_q 
= 
	(-)^{( |A_1| + \cdots + |A_q| ) ( |B_1| + \cdots + |B_q| ) } 
		e^{A_1} \cdots e^{A_p} e^{B_1} \cdots e^{B_q} 
		( \omega_p{}_{A_p\dots A_1} * \omega_q{}_{B_q\dots B_1} ) 
, 
\label{eq:C:Forms} 
\end{equation} 

%%%%%%%%%

The exterior derivative $d$ is defined as the map from a $p$-form to a $p+1$-form
by using the basis $e^A$:
\begin{equation}
d\omega_p=e^{A_1}\cdots e^{A_p} e^{B}D_B\omega_p{}_{A_p\dots A_1}+
 \sum_{r-1}^p (-1)^{|A_{r+1}|+\cdots 
 +|A_p|} e^{A_1}\cdots d e^{A_r} \cdots e^{A_p}\omega_p{}{}_{A_p\dots A_1}
\label{eq:C:derivative}
\end{equation}
with $\omega_p$ in eq.(\ref{eq:C:pForm}) and $d e^A$ is the same as the undeformed one.

Before we start to discuss the Yang-Mills theory 
with the use of the above differential forms, 
we prove the consistency of the deformed exterior algebra.
In our construction the action of the exterior derivative $d$ 
coincides with the undeformed case in the $e^A$-basis 
as seen from eq.\ (\ref{eq:C:derivative}). 
Using eqs.\ (\ref{eq:C:Forms}) and (\ref{eq:C:pForm}),
we can prove the graded Leibniz rule: 
\begin{equation}
d(\omega_p * \omega_q)
=(-1)^{q}d\omega_p * \omega_q+\omega_p * d\omega_q 
. 
\end{equation}
It follows also that $d$ is nilpotent: $d^2=0$.
Finally, the associativity of the deformed exterior algebra 
is a direct consequence of the associativity of the star product.

%%%%%%%%%%%%%%%%%%%
 
Although 
we have used the fact that $e^A$ (anti)commutes with superfields 
to derive eq.(\ref{eq:C:Forms}), 
one should notice that 
general differential forms do not (anti)commute with superfields. 
For example, we have 
\begin{equation} 
Q_\alpha ( dx^\mu ) 
= 
	- i (\sigma^\mu d \bar{\theta})_\alpha 
, 
\end{equation} 
because 
$dx^\mu$ transforms as 
$dx^\mu \rightarrow dx^\mu - i \xi \sigma^\mu d\bar{\theta}$ 
under the supertranslation $\xi^\alpha Q_\alpha$. 
As a result, there are non-trivial commutators 
involving differential forms in $(x, \theta, \bar{\theta})$-coordinates: 
\begin{eqnarray} 
&& 
[ dx^\mu, x^\nu ]_* 
= 
	C^{\alpha\beta} (\sigma^\mu d\bar{\theta})_\alpha 
		(\sigma^\nu \bar{\theta})_\beta 
	, \quad 
[ dx^\mu, \theta^\alpha ]_* 
= 
	i C^{\alpha\beta} (\sigma^\mu d \bar{\theta})_\beta 
	, \quad 
[ d x^\mu ,\bar{\theta}_{\dot{\alpha}} ]_* 
= 
	0 
	, \nonumber\\ 
&& 
[ d \theta^\alpha , f(x, \theta, \bar{\theta}) \}_* 
	= [ d \bar{\theta}_{\dot{\alpha}} , f(x, \theta, \bar{\theta}) \}_* 
	= 0 
, 
\end{eqnarray}
where $f(x, \theta, \bar{\theta})$ is an arbitrary superfield.

\subsection{Reproduction of the ${\cal N}=1/2$ super Yang-Mills theory} 

In the following, we will see that the deformed wedge product defined above 
is consistent with the ${\cal N}=1/2$ super Yang-Mills theory 
described in the previous section, 
in the sense that the curvature 2-from superfield 
will correctly reproduce the field strength superfield 
$W_\alpha$ and $\bar{W}_{\dot{\alpha}}$ in (\ref{eq.Wdeform}) 
(after imposing appropriate constraints as in the undeformed case \cite{GrSoWe}) 
based on the deformed exterior algebra.

Given a connection 1-form superfield $\phi$, 
the curvature superfields $F_{AB}$ are obtained 
as the coefficient functions of 
the two-form superfield $F$ constructed in a standard way: 
\begin{equation} 
F
= 
	d \phi
	+ \phi * \phi 
. 
\label{eq:C:Curvature2-Form}
\end{equation} 
Due to eq.(\ref{eq:C:Forms}), it holds that 
\begin{equation} 
\phi * \phi 
= 
	( e^A \phi_A ) * ( e^B \phi_B ) 
= 
	(-)^{|A| |B|} e^A e^B ( \phi_A * \phi_B ) 
= 
	- \frac12 e^A e^B [ \phi_B ,\phi_A \}_* 
. 
\label{eq:phiphi}
\end{equation} 
Therefore, we find the curvature superfields $F_{AB}$ as 
\begin{equation}
 F_{AB}
=
	D_{A} \phi_{B} - (-)^{|A||B|} D_{B} \phi_{A} 
	-[\phi_A,\phi_B\}_* 
	+T_{AB}{}^{C} \phi_C 
, 
\label{eq:C:Curvature}
\end{equation}
where $T_{AB}{}^C$ is the torsion defined by 
$ 
d e^C 
= 
        \frac12 e^A e^B T_{BA}{}^C 
$ 
whose non-vanishing elements are 
$
T_{\alpha\dot{\beta}}{}^\mu = T_{\dot{\beta}\alpha}{}^\mu 
= 2i \sigma^\mu_{\alpha\dot{\beta}}
$.

The proper constraints for the curvature superfields 
to give the ${\cal N}=1/2$ super Yang-Mills theory 
turn out to be 
\begin{equation}
 F_{\alpha\beta}=0,\ \
 F_{\dot{\alpha}\dot{\beta}}=0,\ \
 F_{\alpha\dot{\beta}}=0, 
\label{eq:C:Constraints} 
\end{equation} 
where the curvature superfields are given by (\ref{eq:C:Curvature}) 
(see \cite{GrSoWe} for the undeformed case). 
We refer these constraints as the Yang-Mills constraints. 
Turning off the deformation, 
these constraints can be solved by 
$ 
\phi_\alpha=-e^{-V} D_\alpha e^V 
$, 
$ 
\phi_{\dot{\alpha}}=0 
$, 
$
\phi_\mu 
= 
	- \frac i4 \bar{\sigma}_\mu^{\dot{\beta}\beta} \bar{D}_{\dot{\beta}} \phi_\beta 
$, 
where $V$ is a general superfield. 
This is checked with the use only of 
the Leibniz rule for the supercovariant derivatives, 
and changing the ordering of the superfields is not needed at all. 
Since the supercovariant derivatives $D_A$ 
satisfy the Leibniz rule even in the presence of the deformation, 
it tells us that 
the spinor connection superfields 
of the same form as in the undeformed case 
are also a solution to the Yang-Mills constraints: 
\begin{equation}
	\phi_\alpha=-e^{-V}_* * D_\alpha e^V_* 
	, \quad 
\phi_{\dot{\alpha}}=0 
	, \quad 
\phi_\mu 
= 
	- \frac i4 \bar{\sigma}_\mu^{\dot{\beta}\beta} \bar{D}_{\dot{\beta}} \phi_\beta 
, 
\label{eq:C:Connections}
\end{equation} 
where $V$ is again a general superfield.

Because of eq.\ (\ref{eq:C:Forms}), we find that 
the curvature superfields $F_{AB}$ satisfy 
the Bianchi identities with the star product: 
\begin{equation} 
\frac12 e^A e^B e^C 
	\left( 
	D_C F_{BA} - [ \phi_C , F_{BA} \}_*  
	+ \frac12 T_{CB}{}^D F_{DA} 
	+ \frac12 T_{CA}{}^D F_{DB} 
	\right) 
= 
	0 
	. 
\label{eq:C:Bianchi} 
\end{equation} 
Determination of all the $F_{AB}$ with the use of the Bianchi identities 
is completely parallel to the undeformed case, 
and we find 
\begin{equation} 
F_{\mu \dot{\alpha}} 
=
      \frac i2 {\cal W}^\beta \sigma_\mu{}_{\beta\dot{\alpha}} 
        , \quad 
F_{\mu \alpha} 
=
      \frac i2 \sigma_\mu{}_{\alpha\dot{\beta}} \bar{\cal W}^{\dot{\beta}} 
\label{eq:F=W} 
\end{equation} 
where ${\cal W}^\alpha = W^\alpha$, 
$\bar{\cal W}^{\dot{\alpha}} = e^{-V}_* * \bar{W}^{\dot{\alpha}} * e^V_*$ 
and $W$ and $\bar{W}$ have the same forms as in the undeformed case 
except for every product replaced with the star product, that is, 
they coincide with the field strength superfields given in (\ref{eq.Wdeform}).

Then the invariant action with respect to super- and gauge symmetry 
can be constructed 
with the use of ${\cal W}$ and $\bar{\cal W}$ as in the undeformed case 
\cite{GrSoWe}
and it is none other than the action $S$ given in (\ref{eq:C:Action}). 
Therefore, imposing the Yang-Mills constraints (\ref{eq:C:Constraints}), 
the ${\cal N}=1/2$ super Yang-Mills theory can be correctly reproduced 
in a geometrical way based on the deformed exterior algebra.

\section{Deformed super ADHM construction}\setcounter{equation}{0} 
\label{sect:DSADHM}

After reviewing the super ADHM construction in section \ref{subsect:Review:SADHM}, 
we describe its non(anti)-commutative deformation 
in section \ref{subsect:DSADHM}. 
The general solution obtained by the deformed construction 
is given in section \ref{subsect:GeneralSolution}.

\subsection{Review of the ${\cal N}=1$ super ADHM construction}
\label{subsect:Review:SADHM} 

In this subsection, 
we briefly review the ${\cal N}=1$ super ADHM construction. 

The ${\rm U}(n)$ (or ${\rm SU}(n)$) $k$ instanton configurations 
can be given by the ADHM construction \cite{AtHiDrMa}. 
Define $\Delta_\alpha (x)$ such as 
\begin{equation} 
\Delta_\alpha (x) 
=
        a_\alpha
        + x_{\alpha\dot{\alpha}} b^{\dot{\alpha}} 
\end{equation} 
where $a_\alpha$ and $b^{\dot{\alpha}}$ are constant 
$k \times(n+2k)$ matrices 
and 
$x_{\alpha\dot{\alpha}} 
\equiv 
        i x_{\mu}\sigma^\mu_{\alpha\dot{\alpha}}
$.
We assume that $\Delta_\alpha$ has 
maximal rank everywhere except for a finite set of points. 
Its hermitian conjugate 
$\Delta^\dagger{}^\alpha \equiv (\Delta_\alpha)^\dagger$ 
is given by 
\begin{equation} 
\Delta^\dagger{}^\alpha (x)
= 
        a^\dagger{}^\alpha 
        + b^\dagger_{\dot{\beta}} x^{\dot{\beta}\alpha} 
. 
\end{equation} 
Then the gauge field $v_\mu$ is given by 
\begin{equation}
v_\mu
=
        - 2 i v^\dagger \partial_\mu v
        \label{eq.binstv}
, 
\end{equation} 
where $v$ is the set of the normalized zero modes of $\Delta_\alpha$: 
\begin{equation} 
\Delta_\alpha v = 0
        , \quad 
v^\dagger v = {\bf 1}_n 
. 
\end{equation}  
For later use we define $f$ which is defined as the inverse of the quantity 
\begin{equation} 
f^{-1} 
\equiv 
        \frac12 \Delta_\alpha \Delta^\dagger{}^\alpha 
. 
\label{eq:finv} 
\end{equation}

The super instanton condition (\ref{eq:SuperInstanton}) can be rewritten 
in the superfield formalism \cite{Sem,Vo} as 
\begin{eqnarray}
&& 
 F_{\mu\dot{\alpha}}=0
\label{eq.SASD} 
        , \\ 
&& 
 \star F_{\mu\nu}=-F_{\mu\nu}
\label{eq.SASD2}
, 
\end{eqnarray} 
where $F$ is the curvature superfield satisfying 
the covariant constraints (\ref{eq:C:Constraints}) and 
the Bianchi identity (\ref{eq:C:Connections}). 
The super ADHM construction gives the solutions 
to the super ASD condition (\ref{eq.SASD}) \cite{ArTaWa}. 
We define 
a superfield extension of $\Delta_\alpha (x)$ 
(see also appendix \ref{app:Notation}):
\begin{equation}
 \hat{\Delta}_\alpha 
                 =\Delta_\alpha (y) + \theta_\alpha {\cal M}
,
\label{eq.SuperDirac}
\end{equation}
where $\Delta_\alpha (y)$ is the zero dimensional Dirac operator in the ordinary
ADHM construction with replacing $x^\mu$ 
by the chiral coordinate $y^\mu = x^\mu + i \theta\sigma^\mu\bar{\theta}$ 
and ${\cal M}$ is a $k\times(n+2k)$ fermionic
matrix which includes the fermionic moduli.
We suppose that $\hat{\Delta}_{\alpha}$ has a maximal rank almost
everywhere as in the ordinary ADHM construction.
Its $\ddagger$-conjugate $\hat\Delta^\ddagger{}^\alpha$ is found to be 
\begin{equation}
\hat{\Delta}^\ddagger{}^{\alpha}
= 
        \Delta^{\dagger}{}^{\alpha} (y) 
        + \theta^\alpha {\cal M}^\dagger 
        . 
\label{def.conjugation}
\end{equation}
As $\hat\Delta_\alpha$ has $n$ zero modes 
we collect them in a matrix superfield $\hat{v}_{[n+2k]\times [n]}$:
\begin{equation}
 \hat{\Delta}_\alpha\hat{v}=0 
. 
\label{eq.Dv}
\end{equation} 
Its $\ddagger$-conjugate $\hat{v}^\ddagger$ satisfies 
$
\hat{v}^\ddagger \hat{\Delta}^\ddagger{}^\alpha = 0
$. 
We require that $\hat{v}$  satisfies the normalization condition: 
\begin{equation}
\hat{v}^\ddagger \hat{v}=1 
. 
\label{eq.normalization}
\end{equation}

The connection one-form superfield $\phi$ is given by 
\begin{equation}
 \phi =- \hat{v}^\ddagger d\hat v.
\label{eq.connectionphi}
\end{equation}
where $d$ is exterior derivative of superspace.
The connection $\phi$ defines the curvature
\begin{equation}
 F 
= 
        d\phi+\phi\phi 
= 
        \hat{v}^\ddagger 
                d \hat{\Delta}^\ddagger{}^{\alpha}\hat K_\alpha{}^{\beta}
        d\hat{\Delta}_\beta\hat{v}
        ,
\label{eq:ADHM:Curvature2-form}
\end{equation}
where
\begin{equation}
\hat K^{-1}{}_\alpha{}^{\beta}
\equiv 
 \hat\Delta_\alpha \hat\Delta^\ddagger{}^{\beta} 
\end{equation}
and $\hat{K}_\alpha{}^\beta$ is defined such that 
$
\hat{K}^{-1}{}_\alpha{}^{\beta} \hat{K}{}_\beta{}^\gamma 
= 
        \hat{K}_\alpha{}^{\beta} \hat{K}^{-1}{}_\beta{}^\gamma  
= 
        \delta_{\alpha}^\gamma {\bf 1}_k 
$. 
Note that we have the following completeness condition:
\begin{equation}
\hat v \hat{v}^\ddagger 
=
	{\bf 1}_{n+2k}
	-\hat\Delta^\ddagger{}^{\alpha} \hat K_\alpha{}^{\beta}\hat\Delta_\beta
.
\end{equation}
The curvature superfield $F_{\mu\nu}$ becomes ASD 
if $\hat K$ satisfies 
$\hat\Delta_\alpha \hat\Delta^\ddagger{}^{\beta}\propto\delta_{\alpha}^{\beta}$ 
and thus 
\begin{equation}
\hat K^{-1}{}_\alpha{}^{\beta} 
=\delta_\alpha^{\beta}\hat{f}^{-1}
\label{eq.diagonal}
\end{equation}
where 
\begin{equation} 
\hat f^{-1} 
\equiv 
        \frac12 \hat{\Delta}_\alpha \hat\Delta^\ddagger{}^\alpha 
\label{eq:fhatinv}
\end{equation} 
is a $k\times k$ matrix superfield.
There exists $\hat f$ because we have assumed that 
$\hat\Delta_\alpha$ has maximal rank. 
The above condition (\ref{eq.diagonal}) leads to 
both the bosonic and fermionic ADHM constraints.  
When eq.\ (\ref{eq.diagonal}) holds, i.e., 
the parameters in $\hat\Delta_\alpha$ are satisfying
both bosonic and fermionic ADHM constraints,
we obtain from eq.(\ref{eq:ADHM:Curvature2-form}) 
the ASD curvature superfield (\ref{eq.SASD2}) 
and another (non-trivial) one 
in terms of the ADHM quantities: 
\begin{eqnarray}
 F_{\mu\nu}
&=& 
        4 \hat{v}^\ddagger b^\dagger \bar{\sigma}_{\mu\nu} \hat{f} b \hat{v} 
\label{eq.adhmfv}
        , \\ 
F_{\mu\alpha} 
&=& 
        \frac i2 \sigma_\mu{}_{\alpha\dot{\beta}} 
        \left\{ 
        - 2 \hat{v}^\ddagger (
                b^\dagger{}^{\dot{\beta}} \hat{f} {\cal M} 
                - {\cal M}^\dagger \hat{f} b^{\dot{\beta}} 
                ) \hat{v} 
   -8 \bar{\theta}_{\dot{\gamma}} 
                \hat{v}^\ddagger ( 
                b^\dagger{}^{\dot{\beta}} \hat{f} b^{\dot{\gamma}} 
                + b^\dagger{}^{\dot{\gamma}} \hat{f} b^{\dot{\beta}} 
                ) \hat{v}
        \right\} 
. 
\end{eqnarray} 
We can check that the other curvature superfields vanish.

To ensure that the superfields obtained by the super ADHM construction 
are correctly in the WZ gauge, 
we impose the following conditions on $\hat v$: 
\begin{equation} 
\bar{D}_{\dot{\alpha}} \hat v 
= 
	0 
	, \quad 
\hat{v}^\ddagger
                {\partial \over \partial \theta^\alpha} \hat{v} 
= 
        0 
. 
\label{eq:v:NecessaryCond} 
\end{equation} 
Imposing these conditions, 
we can determine the zero mode $\hat v$ of $\hat\Delta_\alpha$ as  
\begin{equation} 
\hat{v} 
= 
        v 
        + \theta^\gamma \Bigl( \Delta^\dagger_\gamma f {\cal M} v \Bigr) 
        + \theta\theta 
                \Bigl( \frac12 {\cal M}^\dagger f {\cal M} v \Bigr) 
        , 
\label{eq:vinWZ}
\end{equation} 
and find that the connection superfield $\phi_\mu$ 
constructed as in (\ref{eq.connectionphi}) 
correctly gives the super instanton configuration in the WZ gauge: 
\begin{equation} 
\phi_\mu 
= 
        - \frac i2 \left[ 
        - 2 i v^\dagger \partial_\mu v 
        + i \theta^\gamma \sigma_\mu{}_{\gamma\dot{\beta}} 
                \left\{ 2 i v^\dagger ( b^{\dagger \dot{\beta}} f {\cal M} 
                - {\cal M}^\dagger f b^{\dot{\beta}} ) v \right\} \right]
, 
\end{equation} 
where the lowest component is the instanton gauge field   
and the $\theta$-component is the fermion zero mode.

\subsection{Deformation of the super ADHM construction}
\label{subsect:DSADHM} 

In this subsection, 
extending the super ADHM construction in section \ref{subsect:Review:SADHM}, 
we will present a formulation that provides 
a way to construct deformed ASD instantons 
in the non(anti)-commutative ${\cal N}=1/2$ super Yang-Mills theory,  
i.e. the exact solutions to the deformed equations (\ref{eq.c.sinsteq}).

The deformed super ASD condition turns out to be of the same form 
as the super ASD condition (\ref{eq.SASD}) but the product replaced 
with the star product (\ref{eq:StarProduct1}): 
\begin{eqnarray}
&& 
 F_{\mu\dot{\alpha}}=0
, 
\label{eq:C:SASD} 
	\\ 
&& 
 \star F_{\mu\nu}=-F_{\mu\nu}
, 
\label{eq:C:SASD2}
\end{eqnarray} 
where the curvature superfields $F_{AB}$ are given by eq.(\ref{eq:C:Curvature}). 
Note that 
eq.\ (\ref{eq:C:SASD2}) follows from eq.\ (\ref{eq:C:SASD}) 
as long as the two-form $F$ satisfies 
the Bianchi identities and the Yang-Mills constraints, 
because these imply that 
$
F_{\mu\nu} 
= 
	- \frac14 ( \bar{\cal D} \bar{\sigma}_{\mu\nu} \bar{\cal W} 
	- {\cal D} \sigma_{\mu\nu} {\cal W} ) 
$ 
and also that ${\cal W}$ is proportional to $F_{\mu\dot{\alpha}}$ 
(see eq.\ (\ref{eq:F=W})).

Since the equivalence of the condition (\ref{eq:C:SASD}) and 
the deformed equations (\ref{eq.c.sinsteq}) is not apparent, 
we will show it below. 
We assume that the two-form $F$ satisfies the Bianchi identities 
and the Yang-Mills constraints. 
Then $F_{\mu\dot{\alpha}}$ is proportional to $W_\alpha$ given in (\ref{eq.Wdeform}) 
(or (\ref{eq:C:WComp}) in the WZ gauge) 
as discussed before (see eq.\ (\ref{eq:F=W})). 
If the deformed equations (\ref{eq.c.sinsteq}) are satisfied, 
we immediately find from eq.\ (\ref{eq:C:WComp}) 
that 
$
 W_\alpha=0
$ holds, 
and this implies that eq.\ (\ref{eq:C:SASD}) also holds 
because of eq.\ (\ref{eq:F=W}). 
The converse can be shown as follows. 
First let us assume $\phi_{\dot{\alpha}} = 0$. 
In this case, a solution $\phi_A$ to the Yang-Mills constraints 
can always be written of the form given in (\ref{eq:C:Connections}) 
with $V$ a general superfield. 
Then ${\cal W}$ is determined by eq.\ (\ref{eq:F=W}) and 
will coincide with the field strength superfield $W$ in (\ref{eq.Wdeform}). 
Taking the WZ gauge for $V$ by performing gauge transformations, 
${\cal W}(=W)$ becomes to have the form given in (\ref{eq:C:WComp}) 
provided that the component fields of $V$ are parameterized 
as in (\ref{eq.c.Vwz}). 
Now it is obvious that 
in order for $F_{\mu\dot{\alpha}}$ ($\propto {\cal W}$) to vanish, 
the deformed ASD equations should hold. 
Since all other solutions $\phi_A$ (with $\phi_{\dot{\alpha}} \ne 0$) 
to the Yang-Mills constraints 
are obtained from (\ref{eq:C:Connections}) by gauge transformation 
\begin{equation} 
\phi \rightarrow X^{-1} * \phi * X - X^{-1} * d X 
	, \qquad 
F \rightarrow X^{-1} * F * X 
\end{equation} 
with $X$ an arbitrary invertible (${\rm U}(n)$ valued) superfield, 
the equivalence also holds even 
in the absence of the assumption $\phi_{\dot{\alpha}}=0$.

Let us remark on an implication of the second condition (\ref{eq:C:SASD2}). 
Eq. (\ref{eq:C:SASD2}) 
requires especially that the lowest component of
$F_{\mu\nu}$ contains only the ASD part, 
but this does not mean $v^{\rm SD}_{\mu\nu}=0$, 
since the lowest component is 
not simply $v_{\mu\nu}$ 
but $v_{\mu\nu} + \frac i2 C_{\mu\nu} \bar{\lambda}\bar{\lambda}$ 
in the deformed theory: 
Using eq.(\ref{eq:C:Connections}), 
we find $\phi_\mu$ in the WZ gauge as 
\begin{equation}
\phi_\mu 
= 
        - \frac i2 
        \left[ 
                v_\mu 
                + i \theta \sigma_\mu \bar{\lambda} 
                - \bar{\theta} \bar{\sigma}_\mu W 
        \right] 
        (y) 
. 
\label{eq.c.phiv}
\end{equation}
Here we have used $V_{\mathrm{WZ}}$ given in (\ref{eq.c.Vwz}). 
Then the curvature $F_{\mu\nu}$ is found to be 
\begin{eqnarray} 
F_{\mu\nu}
&=& 
	\partial_\mu\phi_\nu
	- \partial_\nu\phi_\mu
	- [\phi_\mu,\phi_\nu]_*
	\nonumber \\
&=& 
	- \frac{i}{2}
		\left(
		v_{\mu\nu}
		+ \frac{i}{2}C_{\mu\nu}\bar{\lambda}\bar{\lambda} 
		- i \theta\sigma_\mu {\cal D}_\nu\bar{\lambda}
		+ i \theta\sigma_\nu {\cal D}_\mu\bar{\lambda} 
		- i\theta\theta\bar{\lambda}\bar{\sigma}_{\mu\nu}\bar{\lambda}
		\right) 
		\nonumber\\ 
&&{} 
	+ (\mbox{terms containing $\bar{\theta}$ and $W$}) 
	. 
\end{eqnarray} 
Therefore, we can see that 
in order for $F_{\mu\nu}$ to satisfy the ASD condition, 
at least $v^{\rm SD}_{\mu\nu} + \frac i2 C_{\mu\nu} \bar{\lambda}\bar{\lambda}=0$ 
is required. 
In fact, if we use the deformed ASD equations (\ref{eq.c.sinsteq}) 
(therefore $W^\alpha =0$), $F_{\mu\nu}$ reduces to 
\begin{equation} 
F_{\mu\nu} 
= 
	-\frac{i}{2}\left(
                v^{\mathrm{ASD}}_{\mu\nu}
               - i \theta^\alpha \sigma^\rho \bar{\sigma}_{\mu\nu}
			{\cal D}_\rho \bar{\lambda} 
               - i\theta\theta\bar{\lambda}\bar{\sigma}_{\mu\nu}
                                              \bar{\lambda}\right)
	, 
\label{eq.c.fmn}
\end{equation} 
and now it contains only ASD components.

We have seen that 
the deformed ASD equations are equivalent to 
the super ASD condition with the star product, (\ref{eq:C:SASD}). 
One would expect that 
its solutions can be 
constructed by the super ADHM construction, 
replacing each product with the star product (\ref{eq:StarProduct1}). 
In the rest of this subsection, we will see that 
such a deformed ADHM construction actually gives 
solutions to the deformed super ASD condition, 
that is, the deformed super ASD instantons.

For the deformed super ASD instantons, 
$\phi_\mu$ in the WZ gauge becomes 
\begin{equation}
\phi_\mu 
= 
        - \frac i2 
        \left[ 
                v_\mu 
                + i \theta \sigma_\mu \bar{\lambda} 
        \right] 
        (y) 
, 
\end{equation}
since $W_\alpha =0$ holds. 
This again leads us to adopt 
$\hat\Delta_\alpha$  
in our super ADHM construction 
with the same form as in the undeformed case: 
\begin{equation}
 \hat\Delta_\alpha
=
	\Delta_\alpha(y) + \theta_\alpha{\cal M}
. 
\label{eq:C:Delta} 
\end{equation} 
Then, according to the $\ddagger$-conjugation rules, 
we have 
\begin{equation} 
\hat{\Delta}^\ddagger{}^\alpha 
= 
	\Delta^\ddagger{}^\alpha (y) 
	+ \theta^\alpha {\cal M}^\ddagger 
. 
\end{equation} 
Here we will not rewrite $\ddagger$ in the r.h.s.\ with $\dagger$, 
because in the presence of the deformation, 
$\Delta_\alpha$ (and possibly ${\cal M}$) may contain products of Grassmann variables 
and for such quantities we should use 
$\ddagger$ instead of $\dagger$ in general.

We collect the $n$ zero modes of $\hat\Delta$ 
into a matrix form $\hat{u}_{[n+2k]\times[n]}$:
\begin{equation} 
\hat{\Delta}_\alpha * \hat{u} 
= 
	0 
. 
\label{eq:C:Dv} 
\end{equation} 
We require it to be normalized as 
\begin{equation} 
\hat{u}^\ddagger * \hat{u} 
= 
	{\bf 1}_n 
	. 
\label{eq:C:Normalization} 
\end{equation} 
Define $k\times k$ matrices $\hat{K}_*{}_\alpha{}^\beta$ ($\alpha,\beta = 1,2$)
as the ``inverse'' matrices of 
\begin{equation} 
\hat{K}_*^{-1}{}_\alpha{}^\beta 
\equiv 
	\hat{\Delta}_\alpha * \hat{\Delta}^\ddagger{}^\beta 
\end{equation} 
such that 
$
\hat{K}_*^{-1}{}_\alpha{}^{\beta} * \hat{K}_*{}_\beta{}^\gamma 
= 
	\hat{K}_*{}_\alpha{}^{\beta} * \hat{K}_*^{-1}{}_\beta{}^\gamma  
= 
	\delta_{\alpha}^\gamma {\bf 1}_k 
$ 
(see appendix \ref{app:Inverse}). 
Then we have the relation 
\begin{equation} 
\hat{u} * \hat{u}^\ddagger 
= {\bf 1}_{n+2k} - \hat{\Delta}^\ddagger{}^\alpha 
	* \hat{K}_*{}_\alpha{}^\beta * \hat{\Delta}_\beta 
. 
\label{eq:C:Completeness}
\end{equation}

With the use of $\hat{u}$, 
the connection $\phi$ is given by 
\begin{equation}
\phi
= 
	- \hat{u}^\ddagger * d \hat{u} 
. 
\label{eq:C:ADHMConnection}
\end{equation} 

The curvature two-form is given by using the connection one-form $\phi$, 
and now it is written as 
\begin{equation} 
F 
= 
	d\phi + \phi * \phi 
	\nonumber\\ 
= 
	\hat{u}^\ddagger * d \hat{\Delta}^\ddagger{}^\alpha * 
		\hat{K}_*{}_\alpha{}^\beta * d \hat{\Delta}_\beta * \hat{u} 
. 
\end{equation} 
Here we have used (\ref{eq:C:Normalization}), 
(\ref{eq:C:Completeness}) and (\ref{eq:C:Dv}). 
The above equation reads 
\begin{equation}
F_{AB}
=
	- \hat{u}^\ddagger * D_{[A} \hat{\Delta}^\ddagger{}^{\alpha} * 
		\hat{K}_*{}_\alpha{}^{\beta} * D_{B\}} \hat{\Delta}_\beta * \hat{u} 
	. 
\label{eq:C:CurvatureSuperfields}
\end{equation} 
{}From this equation, we find 
\begin{equation} 
F_{\mu\nu} 
= 
	\hat{u}^\ddagger * b^\dagger_{\dot{\alpha}} 
		\bar{\sigma}_{[\mu}{}^{\dot{\alpha}\alpha} \hat{K}_*{}_{\alpha}{}^{\beta}  
		\sigma_{\nu]}{}_{\beta\dot{\beta}} b^{\dot{\beta}} * \hat{u} 
	. 
\end{equation} 
Thus the ASD condition (\ref{eq:C:SASD2}) is satisfied 
if $\hat{K}_*$ commutes with the Pauli matrices: 
\begin{equation} 
\hat{\Delta}_\alpha * \hat{\Delta}^\ddagger{}^\beta 
= 
	\hat{K}_*^{-1}{}_\alpha{}^\beta 
\propto 
	\delta_\alpha^\beta 
. 
\label{eq:C:ADHMConstraint}
\end{equation} 
We immediately find from the expression (\ref{eq:C:CurvatureSuperfields}) that 
$F_{\dot\alpha\dot\beta}=F_{\alpha\dot\beta}=0$ and $F_{\mu\dot\alpha}=0$, 
because $\hat{\Delta}_\alpha$ is a chiral superfield. 
We can also check that 
$F_{\alpha\beta}=0$ with the use of 
the constraint (\ref{eq:C:ADHMConstraint}), 
the relations 
\begin{equation} 
D_\beta \hat{\Delta}_\alpha 
= 
        \varepsilon_{\alpha\beta} 
                ( {\cal M} 
        + 4\bar{\theta}_{\dot{\beta}} b^{\dot{\beta}} ) 
        , \quad 
D_\beta \hat{\Delta}^\ddagger{}^\alpha 
=
        \delta^\alpha_\beta 
                ( {\cal M}^\ddagger 
                + 4 b^\dagger_{\dot{\beta}} 
                        \bar{\theta}^{\dot{\beta}} ) 
, 
\label{eq:DDelta} 
\end{equation} 
and the fact that $F_{\alpha\beta}$ 
is symmetric with respect to $\alpha$ and $\beta$. 
Therefore, we have shown that the above described super ADHM construction gives 
curvature superfields that satisfy 
the Yang-Mills constraints (\ref{eq:C:Constraints}) and 
the ASD conditions (\ref{eq:C:SASD})--(\ref{eq:C:SASD2}) 
if the condition (\ref{eq:C:ADHMConstraint}) is imposed.

The requirement (\ref{eq:C:ADHMConstraint}) 
gives the deformed bosonic and fermionic ADHM constraints as we will see below. 
Because we can write 
\begin{equation} 
\hat{\Delta}_\alpha * \hat{\Delta}^\ddagger{}^\beta
= 
	\hat{\Delta}_\alpha \hat{\Delta}^\ddagger{}^\beta 
	- \frac12 \varepsilon_{\alpha\gamma} C^{\gamma\beta} {\cal M} {\cal M}^\ddagger 
, 
\label{eq:C:DD-C} 
\end{equation} 
we find that the requirement leads to the deformed bosonic ADHM constraint 
\begin{equation} 
\Delta_\alpha \Delta^\ddagger{}^\beta 
	- \frac12 \varepsilon_{\alpha\gamma} C^{\gamma\beta} {\cal M} {\cal M}^\ddagger 
\propto \delta_\alpha^\beta 
, 
\end{equation} 
and the fermionic ADHM constraint 
\begin{equation} 
\Delta_\alpha {\cal M}^\ddagger 
+ {\cal M} \Delta^\ddagger{}_\alpha 
= 
	0 
. 
\end{equation} 
The above equations can also be written as follows (see appendix \ref{app:Notation}): 
\begin{equation}
 a^{\prime\ddagger}_\mu=a_\mu',\ \ \
 \sigma_{\ \beta}^{i\ \ \alpha}\left(
                  a_\alpha a^\ddagger{}^{\beta}
	- \frac12 \varepsilon_{\alpha\gamma} C^{\gamma\beta} {\cal M} {\cal M}^\ddagger 
	\right)=0,
\label{eq.c.badhm}
\end{equation}
\begin{equation}
 {\cal M}'^\ddagger{}^{\dot{\alpha}}={\cal M}^{\prime\dot{\alpha}},\ \ \
 a_\alpha{\cal M}^\ddagger=-{\cal M} a^\ddagger_\alpha. 
\label{eq.c.fadhm}
\end{equation}
These constraints agree with 
those in \cite{BiFrPeLe}  
obtained by considering string amplitudes. 
We can rewrite the deformed ADHM constraints in another form as follows. 
Let us denote 
\begin{equation}
\left( 
 \matrix{ 
  \hat{\Delta}_1 \cr 
  \hat{\Delta}_2 
 } 
\right) 
=
	\left( 
	\matrix{ 
	\hat{J}^\ddagger_{[k]\times[n]} 
		& \bar{z}_2 {\bf 1}_k + \hat{B}_2{}^\ddagger_{[k]\times[k]} 
		& \bar{z}_1 {\bf 1}_k + \hat{B}_1{}^\ddagger_{[k]\times[k]} 
		\cr 
	\hat{I}_{[k]\times[n]} 
		& - z_1 {\bf 1}_k - \hat{B}_1{}_{[k]\times[k]} 
		& z_2 {\bf 1}_k + \hat{B}_2{}_{[k]\times[k]} 
	} 
	\right)  
,
\end{equation}
where $z_1 \equiv y_{2\dot{1}}$, $z_2 \equiv y_{2\dot{2}}$ 
and 
\begin{eqnarray}
&& 
 \hat{I} \equiv I+\theta^1\mu
	, \quad 
 \hat{J} \equiv J+\theta^1\mu^\ddagger
	, \quad 
 \hat{B}_1 \equiv B_1-\theta^1{\cal M}_{\dot{1}}^{\prime}
	, \quad 
 \hat{B}_2 \equiv B_2+\theta^1{\cal M}_{\dot{2}}^{\prime}
	, \nonumber \\ 
&& 
I \equiv \omega_2 
	, \quad 
J^\ddagger \equiv \omega_1 
	, \quad 
B_1 \equiv a'_{2\dot{1}}
	, \quad 
B_2 \equiv a'_{2\dot{2}}
. 
\end{eqnarray} 
Here we have already used 
$a^{\prime\ddagger}_\mu=a_\mu'$ 
and ${\cal M}'^\ddagger{}^{\dot{\alpha}}={\cal M}^{\prime\dot{\alpha}}$. 
Then the constraint (\ref{eq:C:ADHMConstraint}) reads 
\begin{eqnarray}
 \hat{I}*\hat{I}^\ddagger-\hat{J}^\ddagger*\hat{J}+[\hat{B}_1,\hat{B}_1{}^\ddagger]_*
                                    +[\hat{B}_2,\hat{B}_2{}^\ddagger]_*=0 
	, \\
 \hat{I}*\hat{J}+[\hat{B}_2,\hat{B}_1]_*=0
. 
\label{eq.const.sadhm.b}
\end{eqnarray}
In the component language, we find that 
the bosonic ADHM constraints are 
\begin{eqnarray}
 II^\ddagger-J^\ddagger J+[B_1,B_1{}^\ddagger]+[B_2,B_2{}^\ddagger]
	- C^{12}{\cal M}{\cal M}^\ddagger=0
	, \\
 IJ+[B_2,B_1] - \frac{1}{2}C^{11}{\cal M}{\cal M}^\ddagger=0
, 
\end{eqnarray} 
and the fermionic ADHM constraints are
\begin{eqnarray} 
 J^\ddagger\mu^\ddagger - \mu I^\ddagger - [B_1{}^\ddagger,{\cal M}'_{\dot{1}}]
+ [B_2{}^\ddagger,{\cal M}'_{\dot{2}}]=0 
	, \\
 \mu J + I\mu^\ddagger - [B_1,{\cal M}'_{\dot{2}}] 
- [B_2,{\cal M}'_{\dot{1}}]=0.
\end{eqnarray}

After imposing the ADHM constraints, we can write 
$\hat{K}_*^{-1}{}_\alpha{}^\beta 
= \hat{\Delta}_\alpha * \hat{\Delta}^\ddagger{}^\beta
$ as 
\begin{eqnarray} 
&& 
\hat{K}_*^{-1}{}_\alpha{}^\beta 
= 
	\delta_\alpha^\beta \hat{f}^{-1}
, 
\end{eqnarray}
where a $k\times k$ matrix valued superfield $\hat{f}^{-1}$ 
is defined by 
\begin{equation}
\hat{f}^{-1} \equiv \frac12 \hat{\Delta}_\alpha \hat{\Delta}^\ddagger{}^\alpha
, 
\label{eq:C:finv} 
\end{equation}  
which is the same form as in the undeformed case 
since eq.\ (\ref{eq:C:DD-C}) holds. 
As a result, the curvature two-form is written as 
\begin{equation} 
F 
= 
	\hat{u}^\ddagger * d \hat{\Delta}^\ddagger{}^\alpha 
		* \hat{f}_* * d \hat{\Delta}_\alpha * \hat{u} 
, 
\label{eq:C:Fwithf}
\end{equation} 
where $\hat{f}_*$ is defined such that 
$\hat{f}_* * \hat{f}^{-1} = \hat{f}^{-1} * \hat{f}_* = {\bf 1}_k$ 
(see appendix \ref{app:Inverse}). 
Substituting the form of $\hat{\Delta}$, 
we find the following equations from the above expression: 
\begin{eqnarray} 
&& 
F_{\alpha\beta} = F_{\dot{\alpha}\dot{\beta}} = F_{\alpha\dot{\beta}} = 0 
, \\ 
&& 
F_{\mu\nu}
= 
        4 \hat{u}^\ddagger * b^\dagger \bar{\sigma}_{\mu\nu} \hat{f}_* b * \hat{u} 
	, 
	\label{eq:C:FabWithf}
	\\
&& 
F_{\mu\alpha} 
= 
        \frac i2 \sigma_\mu{}_{\alpha\dot{\beta}} 
        \left\{ 
        - 2 \hat{u}^\ddagger * (
                b^\dagger{}^{\dot{\beta}} \hat{f}_* {\cal M} 
                - {\cal M}^\ddagger \hat{f}_* b^{\dot{\beta}} 
                ) * \hat{u} 
	\right. \nonumber\\ 
&&\qquad\qquad \left. {}
   -8 \bar{\theta}_{\dot{\gamma}} 
                \hat{u}^\ddagger * ( 
                b^\dagger{}^{\dot{\beta}} \hat{f}_* b^{\dot{\gamma}} 
                + b^\dagger{}^{\dot{\gamma}} \hat{f}_* b^{\dot{\beta}} 
                ) * \hat{u}
        \right\} 
	, \\ 
&& 
F_{\mu\dot{\alpha}} 
= 
	0 
	. 
\end{eqnarray} 
As mentioned before, the curvature two-form $F$ satisfies 
the Yang-Mills constraints (\ref{eq:C:Constraints}) 
and the deformed super ASD condition (\ref{eq:C:SASD})(\ref{eq:C:SASD2}), 
thus it gives the deformed super ASD instantons.

Owing to the deformed ADHM construction, 
we are able to discuss the dimension of the moduli space
of the deformed instanton solution.
The curvature $F$ is invariant under the following 
${\rm GL}(k) \times {\rm U}(n+2k)$ 
global symmetry transformation 
\begin{equation} 
\hat{\Delta}_\alpha 
\rightarrow 
        G \hat{\Delta}_\alpha \Lambda 
        , \quad 
\hat{f} 
\rightarrow 
        G \hat{f} G^\dagger 
        , \quad 
\hat{v} 
\rightarrow 
        \Lambda^{-1} \hat{v} 
, 
\end{equation} 
where $G \in {\rm GL}(k)$ and $\Lambda \in {\rm U}(n+2k)$. 
Note that we cannot take $G$ and $\Lambda$ as superfields, 
since the form of $\hat{\Delta}_\alpha$ is significant 
for the two-form $F$ (\ref{eq:C:CurvatureSuperfields}) to satisfy 
the Yang-Mills constraints as well as the deformed super ASD condition. 
After fixing $b$ in the canonical form (see (\ref{eq:b:Canonical})), 
the global symmetry breaks down to ${\rm U}(n) \times {\rm U}(k)$ 
as in the purely bosonic ADHM construction, 
and the ${\rm U}(n)$ transformation is considered as a part of the gauge transformation. 
Therefore, the number of the bosonic moduli contained 
in $\hat \Delta_\alpha$ is 
$4nk$ after imposing the 
bosonic ADHM constraints (\ref{eq.c.badhm}) 
and modding out by the ${\rm U}(k)$ symmetry, 
as in the undeformed case. 
There is no additional symmetry and the number of fermionic parameters is reduced 
simply by the fermionic ADHM constraints (\ref{eq.c.fadhm}) 
and we have $2kn$ fermionic moduli as in the undeformed case.

\subsection{The general solution in the Wess-Zumino gauge} 
\label{subsect:GeneralSolution} 

In this subsection, 
we give an expression in terms of the ADHM data $\Delta_\alpha$ and ${\cal M}$, 
of the general solution in the WZ gauge obtained by our construction.

In the WZ gauge, $F_{\mu\nu}$ is a chiral superfield 
because $\phi_\mu$ is so. 
Since we are interested in the field strength in the WZ gauge, 
we find from the expression (\ref{eq:C:FabWithf}) that 
it is sufficient to restrict the zero mode $\hat{u}$ to a chiral superfield. 
Hereafter we restrict the zero mode $\hat{u}$ to a chiral superfield, 
which we write as  
\begin{equation} 
\hat{u} 
= 
	u^{(0)} 
	+ \theta^\gamma u^{(1)}_\gamma 
	+ \theta\theta u^{(2)} 
. 
\end{equation}

When $\hat{u}$ and $\hat{u}^\ddagger$ are chiral superfields, 
we find from (\ref{eq:C:ADHMConnection}) that 
$
\phi^{\dot{\alpha}}
= 
	0 
, 
$
and that $\phi_\mu$ is a chiral superfield 
($
\bar{D}_{\dot{\alpha}} \phi_\mu = 0 
$) 
because $\phi_\mu$ is given by 
\begin{equation} 
\phi_\mu 
= 
	- \hat{u}^\ddagger  
		* {\partial \over \partial y^\mu} \hat{u} (y, \theta) 
. 
\end{equation} 
These are consistent with the connection superfields for the super instantons. 
The rest of the connection $\phi_{\alpha}$ gives a non-trivial necessary condition. 
In the ADHM construction, $\phi_\alpha$ can be written in the chiral basis as  
\begin{equation} 
\phi_\alpha 
=
	- \hat{u}^\ddagger * D_\alpha \hat{u} 
= 
	- \hat{u}^\ddagger 
		* {\partial \over \partial \theta^\alpha} \hat{u} (y, \theta) 
	- 2 i (\sigma^\mu \bar{\theta})_{\alpha} \hat{u}^\ddagger 
		* {\partial \over \partial y^\mu } \hat{u} (y, \theta)
. 
\end{equation} 
We should notice that only the first term is $\bar{\theta}$-independent. 
In the WZ gauge, $\phi_\alpha$ is given by 
\begin{equation} 
\phi_\alpha 
= 
	- e_*^{-V_{\rm WZ}} * D_\alpha e_*^{V_{\rm WZ}} 
= 
	- D_\alpha V_{\rm WZ} + \frac12 [ V_{\rm WZ} , D_\alpha V_{\rm WZ} ]_* 
. 
\end{equation} 
Because $V_{\rm WZ}$ contains at least one $\bar{\theta}$ in each term, 
$\phi_\alpha$ should not have any $\bar{\theta}$-independent terms. 
As a result, it should hold that 
\begin{equation} 
\hat{u}^\ddagger 
		* {\partial \over \partial \theta^\alpha} \hat{u} (y, \theta) 
= 
	0 
	, 
\label{eq:C:Necessary} 
\end{equation} 
which is a necessary condition for $\hat{u}$ to be in the WZ gauge. 
We can use this condition to determine $\hat{u}$ in the WZ gauge.

For convenience, let us define a $k\times k$ matrix 
\begin{equation} 
K^{-1}{}_\alpha{}^\beta 
\equiv 
	\Delta_\alpha \Delta^\ddagger{}^\beta 
\end{equation} 
and its inverse $K$ such that 
$ 
K_\alpha{}^\gamma K^{-1}{}_\gamma{}^\beta 
= K^{-1}{}_\alpha{}^\gamma K_\gamma{}^\beta 
= \delta_\alpha^\beta {\bf 1}_k 
$. 
After imposing the ADHM constraint, we have 
\begin{equation} 
K^{-1}{}_\alpha{}^\beta 
=
	\delta_\alpha^\beta f^{-1} 
	+ \frac12 \varepsilon_{\alpha\gamma} C^{\gamma\beta}\mathcal{M}\mathcal{M}^\ddagger
	, \quad 
f^{-1} 
\equiv 
	\frac 12 \Delta_\gamma \Delta^\ddagger{}^\gamma 
. 
\end{equation} 
where $f^{-1}$ is defined as the lowest component of $\hat{f}^{-1}$ 
in (\ref{eq:C:finv}). 
Defining a matrix 
\begin{equation} 
{\cal C} 
=
	\left(\matrix{ {\cal C}_1{}^1 & {\cal C}_1{}^2 
	\cr {\cal C}_2{}^1 & {\cal C}_2{}^2 } \right) 
	, \quad 
{\cal C}_\alpha{}^\beta 
\equiv 
	\frac12 \varepsilon_{\alpha\gamma} C^{\gamma\beta}
, 
\end{equation} 
we can write 
$K^{-1}{}_\alpha{}^\beta 
= ( {\bf 1}_2 \otimes {\bf 1}_n + {\cal C} \otimes\mathcal{M}\mathcal{M}^\ddagger f 
)_\alpha{}^\beta f^{-1}$ 
where $f$ is the inverse of $f^{-1}$ such that 
$ 
f f^{-1} = f^{-1} f = {\bf 1}_k
$. 
Then we find an expression of the matrix $K$: 
\begin{eqnarray} 
K_\alpha{}^\beta 
= 
	f 
	\Bigl( {\bf 1}_n + \det {\cal C} (\mathcal{M}\mathcal{M}^\ddagger f )^2 \Bigr)^{-1} 
	\left\{ 
	{\bf 1}_2 \otimes {\bf 1}_n 
	- {\cal C} \otimes\mathcal{M}\mathcal{M}^\ddagger f 
	\right\}{}_\alpha{}^\beta 
. 
\label{eq:C:K} 
\end{eqnarray} 
Here we have used a relation 
${\cal C}_\alpha{}^\gamma {\cal C}_\gamma{}^\beta 
= - \delta_\alpha^\beta \det {\cal C}$. 
\footnote{
{}From this expression we find the following useful relations: 
$$
K_\alpha{}^\beta {\cal C}_\beta{}^\gamma 
= 
	{\cal C}_\alpha{}^\beta K_\beta{}^\gamma 
	, \quad 
{\cal C}_\alpha{}^\gamma K_\gamma{}^\delta {\cal C}_\delta{}^\beta 
= 
	- \det {\cal C} K_\alpha{}^\beta 
. 
$$
}

With a given $v$ that satisfies 
$\Delta_\alpha v = 0$ and $v^\ddagger v ={\bf 1}_n$, 
we can determine the zero mode superfield $\hat{u}$ 
: 
\begin{equation} 
\hat{u} 
= 
	\Bigl( 
	{\bf 1}_{n+2k} 
	+ \theta^\gamma \Delta^\ddagger{}_\gamma f\mathcal{M}
	+ \theta \theta \frac12 \mathcal{M}^\ddagger f\mathcal{M}
	\Bigr) u^{(0)} 
, 
\label{eq:C:ZeroMode:Gen} 
\end{equation} 
where 
\begin{equation} 
u^{(0)} 
= 
	\Bigl\{ 
	v 
	- \frac12 ( \Delta^\ddagger K \varepsilon C \Delta ) 
		\mathcal{M}^\ddagger Z^{-1} f\mathcal{M}v 
	\Bigr\} N^{-1/2} U  
, 
\label{eq:C:ZeroMode0} 
\end{equation} 
$U$ is a unitary matrix such that $U^\dagger = U^{-1}$, 
$K$ is given by (\ref{eq:C:K}) 
and $Z,N$ are 
\begin{eqnarray} 
Z 
&\equiv& 
	{\bf 1}_{k} 
	+ \frac12 f\mathcal{M}( \Delta^\ddagger K \varepsilon C \Delta ) \mathcal{M}^\ddagger 
	, 
\label{eq:C:Z} 
	\\ 
N 
&\equiv& 
	{\bf 1}_n 
	+ \frac14 \det C v^\ddagger \mathcal{M}^\ddagger f Z^{-1}{}^\ddagger
		\mathcal{M} v v^\ddagger \mathcal{M}^\ddagger Z^{-1} f\mathcal{M}v 
. 
\label{eq:C:N} 
\end{eqnarray} 
Readers can check that the above $\hat{u}$ is indeed 
a normalized zero mode of $\hat{\Delta}_\alpha$, 
satisfying the WZ gauge condition (\ref{eq:C:Necessary}) 
(for the detailed derivation, see appendix \ref{app:Determination}). 
We have used the normalized zero mode $v$ of $\Delta_\alpha$ 
and it will be found as follows. 
If we solve the deformed ADHM constraints, 
the zero dimensional Dirac operator $\Delta_\alpha$ (the fermionic moduli ${\cal M}$) 
can be separated 
into the $C$-independent part $\Delta_{0}{}_\alpha$ (${\cal M}_0$) and 
the residual $C$-dependent part $\delta \Delta_\alpha$ ($\delta {\cal M}$): 
\begin{equation}
 \Delta_\alpha
  =\Delta_0{}_\alpha+\delta\Delta_\alpha 
	, \quad 
 \mathcal{M}=\mathcal{M}_0{}+\delta\mathcal{M}
. 
\label{eq:C:Split} 
\end{equation}
Then the zero mode $v$ of $\Delta_\alpha$ is given by 
\begin{eqnarray}
v
&=& 
	(
	{\bf 1}_{n+2k} 
	+ \Delta_{0}^{\dagger\alpha} f_{0}\delta\Delta_\alpha
	)^{-1} v_{0} 
	\nonumber\\ 
&&{} 
	\times \left\{ v_0^\dagger 
		( {\bf 1}_{n+2k} 
			+ \delta \Delta^\ddagger{}^\gamma f_0 \Delta_0{}_\gamma )^{-1} 
		( {\bf 1}_{n+2k} 
			+ \Delta_0^\dagger{}^\gamma f_0 \delta \Delta_\gamma )^{-1} v_0 
	\right\}^{-1/2} 
, 
\label{eq.c.vprime}  
\end{eqnarray}
where $v_{0}$ satisfies $\Delta_{0}{}_\alpha v_{0} = 0$ 
as well as the completeness relation 
$ 
v_{0} v_{0}^\dagger 
=
	{\bf 1}_{n+2k}
	- \Delta_{0}^{\dagger\alpha} f_{0}\Delta_{0}{}_\alpha
$. 
Here 
$f_{0}^{-1} 
\equiv 
	\frac12 \Delta_{0}{}_\gamma \Delta_{0}^{\dagger\gamma} 
$ 
and $f_{0}$ is its inverse matrix. 
We can check that the above $\hat{u}$ satisfies simultaneously 
the zero mode equation (\ref{eq:C:Dv}), 
the normalization condition 
(\ref{eq:C:Normalization}) 
and the WZ gauge condition (\ref{eq:C:Necessary}).

The connection superfield $\phi_\mu$ is constructed 
with the use of $\hat{u}$ in (\ref{eq:C:ZeroMode:Gen}): 
\begin{eqnarray}
\phi_\mu
&=& 
	- u^{(0)\ddagger}\partial_\mu u^{(0)}
	\nonumber\\ 
&&{} 
	+ \frac{1}{2}u^{(0)\ddagger}\mathcal{M}^\ddagger{} f
      (\varepsilon C)_\beta{}^\alpha\Delta_\alpha\partial_\mu\Delta^\ddagger{}^\beta
         f\mathcal{M} u^{(0)} 
	- \frac{1}{4}\det Cu^{(0)\ddagger}\mathcal{M}^\ddagger{} 
		f\mathcal{M}\mathcal{M}^\ddagger{}
      \partial_\mu(f\mathcal{M} u^{(0)}) 
	\nonumber\\ 
&&{} 
	- \theta^\alpha u^{(0)\ddagger}(\partial_\mu\Delta^\ddagger{}_\alpha f\mathcal{M}
          +\mathcal{M}^\ddagger{} f\partial_\mu\Delta_\alpha)u^{(0)} 
.
\label{eq.c.phivv0}
\end{eqnarray}
Here we have used the following relations: 
\begin{eqnarray} 
&& 
 \Delta_\alpha u^{(0)}
  +\frac{1}{2}(\varepsilon C)_\alpha{}^\beta\Delta_\beta\mathcal{M}^\ddagger{}
      f\mathcal{M} u^{(0)}=0
	, \\ 
&& 
 \partial_\mu\Delta_\alpha\mathcal{M}^\ddagger{}
  +\mathcal{M}\partial_\mu\Delta^\ddagger{}_\alpha=0
	, \\ 
&& 
\partial_\mu f 
= 
	f \Delta^\gamma \partial_\mu \Delta^\ddagger{}_\gamma f 
. 
\label{eq:Delf} 
\end{eqnarray} 
The first equation can be shown with the use of 
eq.\ (\ref{eq:C:ZeroMode0}) and 
$$ 
Z^{-1} 
= 
	{\bf 1}_{k} 
	- \frac12 f {\cal M} ( \Delta^\ddagger K \varepsilon C \Delta ) 
		{\cal M}^\ddagger Z^{-1} 
. 
$$ 
The last equation follows from the bosonic ADHM constraints. 

Now, because of the WZ gauge, we are able to compare our connection one-form  
to the solutions obtained in the component formalism. 
In appendix \ref{app:U(2)OneInstanton}, 
we show that the known ${\rm U}(2)$ one instanton solution 
is obtained by our construction.

\section{Conclusions and discussion}\setcounter{equation}{0}
\label{sect:Conclusion}

In this paper, we have extended the super ADHM construction to give 
solutions to the deformed ASD instanton equations 
in ${\cal N}=1/2$ super Yang-Mills theory with ${\rm U}(n)$ gauge group. 

First we have extended the exterior algebra on superspace 
to non(anti)commutative superspace, 
and shown that it is a consistent deformation such that 
the field strength superfields of ${\cal N}=1/2$ super Yang-Mills theory 
are correctly reproduced by a curvature two-form superfield. 
We found 
the covariant constraints on the curvature two-form 
(referred to as Yang-Mills constraints) 
and the super ASD condition for the deformed ASD instantons. 

Based on the deformed exterior algebra, 
we have formulated a non(anti)commutative version of the ADHM construction 
and shown that the resulting curvature two-form superfield indeed satisfies 
the Yang-Mills constraints as well as the super ASD condition. 
This means that our construction correctly gives 
deformed ASD instantons 
in the ${\cal N}=1/2$ super Yang-Mills theory. 
We have seen that deformation terms emerge in the bosonic ADHM constraints 
(see also \cite{BiFrPeLe}), 
which are comparable with the ${\rm U}(1)$ terms 
due to space-space noncommutativity \cite{NeSc}. 
Our formulation reveals the geometrical meaning of 
those deformation terms as non(anti)commutativity of superspace.

The deformed super ADHM construction will facilitate us 
to discuss the moduli space of the deformed ASD instantons 
in the ${\cal N}=1/2$ super Yang-Mills theory. 
In this paper, we saw 
that the numbers of the bosonic and fermionic moduli parameters 
of our solutions are the same as the ordinary theory: 
They are $4kn$ and $2kn$ respectively, where $k$ is the instanton number. 
Additional moduli parameters, if they exist, 
may be contained in the $\theta\theta$-component of $\hat{\Delta}_\alpha$, 
but this would change, for example, the ADHM constraints, 
leading to a discrepancy with the result in \cite{BiFrPeLe}.  
We believe that our construction gives all the deformed ASD instantons, 
but it would be interesting to check it directly 
by considering reciprocity \cite{CoGo}. 

Finally we would like to give 
a comment on a relation between the deformed ADHM constraints 
and the hyper-K\"ahler quotient construction \cite{HiKaLiRo}. 
In the ordinary (commutative or noncommutative) gauge theory, 
the fermionic ADHM constraints ensure that 
the fermionic moduli 
are Grassmann-valued symplectic tangent vectors of the bosonic moduli space. 
Since the fermionic ADHM constraints 
are not modified in the present case, 
we expect that this interpretation is not modified. 
On the other hand, our bosonic ADHM constraints contain 
deformation terms which are $k\times k$ matrices, 
not just ${\rm U}(1)$ terms in general.
If there is a ${\rm U}(1)$ term, 
it is well known that setting a particular value of the term 
corresponds to choosing a particular level set 
in the $n+2k$ dimensional mother space 
in the hyper-K\"ahler quotient construction. 
The ordinary instantons 
or the localized instantons \cite{Fu}  
correspond to this value set to be zero, 
and the Nekrasov-Schwarz instantons \cite{NeSc}  
correspond to 
this value set to be a non-zero constant. 
It needs a further study to clarify  
how our deformation terms can be interpreted 
in the hyper-K\"ahler quotient construction.

{\bf Acknowledgments}:
T.~A. is very grateful to the members of 
the Particle Theory and Cosmology Group 
at Tohoku University for their support and hospitality.
This research is partly supported by Grant-in-Aid for Scientific
Research from the Ministry of Education, Culture,
Sports, Science and Technology, Japan No.\ 13640256 and 13135202. 
T.~A. is supported by 
Grant-in-Aid for Scientific Research in Priority Areas (No.\ 14046201) 
from the Ministry of Education, Culture, Sports, Science 
and Technology.

\renewcommand{\theequation}{\Alph{section}.\arabic{equation}}
\appendix 

\section{Notation and Conventions}\setcounter{equation}{0} 
\label{app:Notation} 

We use the following sigma matrices: 
\begin{equation}
\sigma^\mu = \sigma_\mu
\equiv 
        (-i{\bf 1}, \sigma^i)
        , \quad 
\bar{\sigma}^\mu = \bar{\sigma}_\mu
\equiv 
        (-i{\bf 1}, -\sigma^i)
, 
\end{equation}
where $\sigma^i$ are the Pauli matrices and 
$
\bar{\sigma}^{\mu}{}^{\dot{\alpha}\alpha}
= 
        \varepsilon^{\dot{\alpha}\dot{\beta}}
        \varepsilon^{\alpha\beta}\sigma_{\beta\dot{\beta}}^\mu
$ 
holds. 
The Lorentz generators are 
\begin{equation}
\sigma^{\mu\nu} 
\equiv  
        \frac14 (\sigma^\mu \bar{\sigma}^\nu - \sigma^\nu \bar{\sigma}^\mu) 
, \quad 
\bar{\sigma}_{\mu\nu} 
\equiv \frac14 ( \bar{\sigma}_\mu \sigma_\nu - \bar{\sigma}_\nu \sigma_\mu )
, 
\end{equation} 
where 
\begin{equation} 
\sigma^{\mu\nu} 
= 
        \frac {1}{2} \varepsilon^{\mu\nu\lambda\rho} 
                \sigma_{\lambda\rho} 
, \quad 
\bar{\sigma}^{\mu\nu} 
= 
        - \frac {1}{2} \varepsilon^{\mu\nu\lambda\rho} 
                \bar{\sigma}_{\lambda\rho} 
, \quad 
\varepsilon^{0123} = \varepsilon_{0123} \equiv -1  
. 
\end{equation}
They can be  written as
\begin{equation}
 \sigma^{\mu\nu}=-\frac{i}{2}\eta^i_{\mu\nu}\sigma^i,
  \quad
 \bar{\sigma}^{\mu\nu}=-\frac{i}{2}\bar{\eta}^i_{\mu\nu}\sigma^i
\end{equation}
in terms of 't Hoot's eta symbol $\eta^i_{\mu\nu}$, $\bar{\eta}^i_{\mu\nu}$.
The 't Hoot's eta symbol is defined by
\begin{equation}
  \eta^i_{\mu\nu}\equiv(\epsilon_{i\mu\nu0}
  -\delta_{\mu 0}\delta_{i\nu}+\delta_{\nu 0}\delta_{i\mu}),
  \quad
    \bar{\eta}^i_{\mu\nu}\equiv(\epsilon_{i\mu\nu0}
          +\delta_{\mu 0}\delta_{i\nu}-\delta_{\nu 0}\delta_{i\mu})
\end{equation}

We define 
\begin{equation} 
x_{\alpha\dot{\beta}}
\equiv 
        i x_\mu \sigma^\mu_{\alpha\dot{\beta}} 
        , \quad 
x^{\dot{\beta}\alpha}  
\equiv 
        i x_\mu \bar{\sigma}_\mu^{\dot{\beta}\alpha} 
. 
\end{equation}

The ``zero dimensional Dirac operator'' in the extended ADHM construction 
is defined by 
\begin{equation} 
\hat{\Delta}_\alpha
        {}_{[k]\times[n+2k]} 
= 
        \Delta_\alpha (y) 
        + \theta_\alpha {\cal M} 
        , 
\end{equation} 
where 
\begin{equation} 
\Delta_\alpha (x) 
\equiv 
        a_\alpha
        + x_{\alpha\dot{\beta}} b^{\dot{\beta}}
\end{equation} 
and 
\begin{eqnarray}
&& 
a_{\alpha}
        {}_{[k]\times[n+2k]}  
\equiv 
        \Bigl(\matrix{
                \omega_{\alpha}
                        {}_{[k]\times[n]}
                & 
                (a'{}_{\alpha\dot{\beta}})
                        {}_{[k]\times[2k]}
        }\Bigr) 
= 
        \Bigl(\matrix{
                (\omega_{\alpha}{}^i{}_u) 
                & 
                (a'{}_{\alpha\dot{1}}{}^i{}_j) 
                & 
                (a'{}_{\alpha\dot{2}}{}^i{}_j)
        }\Bigr) 
        , \nonumber\\
&& 
{\cal M} 
        {}_{[k]\times[n+2k]}  
\equiv 
        \Bigl(\matrix{
                \mu
                        {}_{[k]\times[n]}
                & 
                ({\cal M}'{}_{\dot{\beta}}) 
                        {}_{[k]\times[2k]}
        }\Bigr) 
= 
        \Bigl(\matrix{
                (\mu^i{}_u) 
                & 
                ({\cal M}'{}_{\dot{1}}{}^i{}_j) 
                & 
                ({\cal M}'{}_{\dot{2}}{}^i{}_j) 
        }\Bigr) 
,
\label{def.app.moduli}
\end{eqnarray} 
with $u=1,\dots ,n$ and $i,j=1,\dots , k$. 
Note that we write 
$a'{}_{\alpha\dot{\beta}}{}^j{}_i 
= 
        i a'^\mu{}^j{}_i \sigma_\mu{}_{\alpha\dot{\beta}}
$. 
The canonical form of $b$ is defined as 
\begin{equation}
(b^{\dot{\alpha}})
= 
        \left(\matrix{
        b^{\dot{1}} \cr b^{\dot{2}}
        }\right)
= 
        \left(\matrix{
        {\bf 0}_{[k]\times [n]} & {\bf 1}_k & {\bf 0}_k \cr 
        {\bf 0}_{[k]\times [n]} & {\bf 0}_k & {\bf 1}_k 
        }\right) 
. 
\label{eq:b:Canonical} 
\end{equation}

\section{The inverse superfield}\setcounter{equation}{0}
\label{app:Inverse} 

In this appendix, 
we give an expression of the ``inverse'' of a chiral superfield. 
Assume $(\phi_1{}^i{}_j)$ is an invertible $k\times k$ matrix. 
For a $k\times k$ matrix valued superfield 
\begin{equation} 
\Phi(y,\theta) = \phi_1(y) + \sqrt{2} \theta \psi_1(y) + \theta\theta F_1(y), 
\end{equation} 
its ``inverse'' superfield $\Phi^{-1}(y,\theta)$ is defined by the relation 
\begin{equation} 
\Phi * \Phi^{-1} = \Phi^{-1} * \Phi = {\bf 1}_k
. 
\end{equation} 
Explicitly it is given by 
\begin{equation} 
\Phi^{-1}(y,\theta) = \phi_2(y) + \sqrt{2} \theta \psi_2(y) + \theta\theta F_2(y), 
\end{equation}
where 
\begin{eqnarray} 
\phi_2 
&=& 
	( {\cal A} + \det C {\cal F} {\cal A}^{-1} {\cal F} )^{-1} 
	, \\
\psi_2{}_\alpha 
&=& 
	- {\cal A}'^{-1} 
	( 
	\Psi_\alpha 
	+ \varepsilon_{\alpha\beta} C^{\beta\gamma} \Psi_\gamma 
		{\cal A}^{-1} {\cal F} 
	) 
	\phi_2 
	, \\
F_2 
&=& 
	- {\cal A}^{-1} {\cal F} \phi_2
. 
\end{eqnarray}
Here we have defined the following quantities: 
\begin{eqnarray} 
{\cal A}' 
&\equiv& 
	\phi_1  
	+ \det C F_1 \phi_1^{-1} F_1 
	, \\
\Psi_\alpha 
&\equiv& 
	( 
	\delta_\alpha^\beta 
	- \varepsilon _{\alpha\gamma} C^{\gamma\beta} F_1 \phi_1^{-1} 
	) 
	\psi_1{}_\beta 
	, \\
{\cal A} 
&\equiv& 
	\phi_1 
	+ C^{\alpha\beta} \psi_1{}_\alpha {\cal A}'^{-1} \Psi_\beta 
	, \\
{\cal F} 
&\equiv& 
	F_1 
	+ \psi_1^\alpha {\cal A}'^{-1} \Psi_\alpha 
	. 
\end{eqnarray}

\section{Determination of the zero mode superfield}\setcounter{equation}{0} 
\label{app:Determination}

In this appendix, we give a detail of determination of 
the zero mode superfield $\hat{u}$ 
such that it correctly satisfies 
the normalization condition (\ref{eq:C:Normalization}) 
and the WZ gauge conditions (\ref{eq:C:Necessary}).

Let us begin with solving the zero mode eq. (\ref{eq:C:Dv}).
This equation reads 
\begin{eqnarray} 
0 
&=& 
	\Delta_\alpha u^{(0)} 
	- \frac12\mathcal{M}( \varepsilon C u^{(1)} )_\alpha 
	\nonumber\\ 
&&{} 
	+ \theta^\beta 
		\left( 
		\Delta_\alpha u^{(1)}_\beta 
		+ \varepsilon_{\alpha\beta}\mathcal{M}u^{(0)} 
		- C_{\alpha\beta}\mathcal{M}u^{(2)} 
		\right) 
	\nonumber\\ 
&&{} 
	+ \theta\theta 
		\left( 
		\Delta_\alpha u^{(2)} 
		+ \frac12\mathcal{M}u^{(1)}_\alpha 
		\right)
\label{eq:app:C:ZeroMode:Comp} 
\end{eqnarray} 
where 
$( \varepsilon C )_\alpha{}^\beta 
\equiv \varepsilon_{\alpha\gamma} C^{\gamma\beta}$. 
With a given $v$ that satisfies 
$\Delta_\alpha v = 0$ and $v^\ddagger v ={\bf 1}_n$, 
$u^{(0)}$ and $u^{(2)}$ can be written with the use of $u^{(1)}_\alpha$ as 
\begin{eqnarray} 
&& 
u^{(0)} 
= 
	\frac12 ( \Delta^\ddagger K )^\gamma\mathcal{M}( \varepsilon C u^{(1)} )_\gamma 
	+ v c 
\label{eq.c.v0}
	, \\ 
&& 
u^{(2)} 
= 
	- \frac12 ( \Delta^\ddagger K )^\gamma\mathcal{M}u^{(1)}_\gamma 
	+ v t
\label{eq.c.v2}
, 
\end{eqnarray}
where 
$( \Delta^\ddagger K )^\alpha 
\equiv \Delta^\ddagger{}^\gamma K_\gamma{}^\alpha$ 
and $c,t$ are arbitrary $n \times n$ bosonic matrices. 
Substituting (\ref{eq.c.v0}) and (\ref{eq.c.v2}) into
(\ref{eq:app:C:ZeroMode:Comp}),
we find that the zero mode equation (\ref{eq:app:C:ZeroMode:Comp})
becomes
\begin{equation}
 \Delta_\alpha u^{(1)}{}^\beta
  -\mathcal{M} u(\delta_\alpha{}^{\beta}c
  +(\varepsilon C)_\alpha{}^{\beta} t )
  -\frac{1}{2}\mathcal{M}(\Delta^\ddagger K)^{\gamma'}\mathcal{M}
     (\delta_\alpha{}^{\beta}(\varepsilon C)_{\gamma'}{}^{\gamma}
        -(\varepsilon C)_\alpha{}^{\beta}\delta_{\gamma'}{}^{\gamma})u^{(1)}_{\gamma}=0.
\label{eq.eqv1}
\end{equation}

We will solve this equation.
The $n+2k$ dimensional space is spanned by the $n+2k$ column vectors
$\{\Delta^\ddagger{}^\alpha,v\}$,
thus in general, we can write $u^{(1)}{}^\alpha$ as
\begin{equation}
 u^{(1)}{}^\alpha
  =v\gamma^\alpha
   +\Delta^\ddagger{}^\beta r_\beta{}^\alpha.
\label{eq.r_gen} 
\end{equation}
where $\gamma^\alpha$ is a $k\times k$ fermionic matrix,
and $r_\beta{}^\alpha$ is $k \times n$ one.
As eq.\ (\ref{eq.eqv1}) is composed of four independent equations 
with respect to the spinor indices, 
we can split it into one proportional to $\delta_\alpha{}^\beta$,
one proportional to $(\varepsilon C)_\alpha{}^\beta$ 
and the other two equations. 
Substituting the general form (\ref{eq.r_gen}) into (\ref{eq.eqv1}), 
we find that the latter two equations are
\begin{equation}
   f^{-1}r'{}_\alpha{}^\beta
      +\frac{1}{2}(\varepsilon C)_\alpha{}^\gamma 
	r'{}_\gamma{}^\beta\mathcal{M}\mathcal{M}^\ddagger
   =0, 
\end{equation}
where $r'{}_\alpha{}^\beta$ represents the terms contained in $r{}_\alpha{}^\beta$ 
which are not proportional to $\delta_\alpha{}^\beta$ 
or $(\varepsilon C)_\alpha{}^\beta$. 
Because of the explicit $C$-dependence of the second term in the l.h.s., 
we find that $r'{}_\alpha{}^\beta$ should vanish 
by a perturbative argument with respect to $C$
\footnote{We assume that the deformation parameter dependence 
admits a perturbative expansion. }
. 
Thus the form of $u^{(1)}{}^\alpha$ is simplified as
\begin{equation}
 u^{(1)}{}^\alpha
  =v\gamma^\alpha
   +\Delta^\ddagger{}^\alpha r
   +\Delta^\ddagger{}^\beta (\varepsilon C)_\beta{}^\alpha s
\label{eq.v1rs},
\end{equation}
where $r$ and $s$ are $k \times n$ fermionic matrices.
Substituting this equation into eq.\ (\ref{eq.eqv1}),
$r$ and $s$ are written in terms of $c$, $t$ and $\gamma^\alpha$.
As a result, the zero mode equation (\ref{eq:app:C:ZeroMode:Comp}) is
solved by (\ref{eq.c.v0}), (\ref{eq.c.v2}) and
(\ref{eq.eqv1}) with
\begin{eqnarray}
 \left(\matrix{
  r\cr
  s
 }\right)
&=&Z^{-1}f(1+\frac{1}{4}\det C
         (\mathcal{M} v v^\ddagger \mathcal{M}^\ddagger Z^{-1}f)^2)^{-1}
         \nonumber\\
 && \times 
\left(\matrix{
 1&
  \frac{1}{2}\det C \mathcal{M} v v^\ddagger \mathcal{M}^\ddagger Z^{-1}f\cr
 -\frac{1}{2}\mathcal{M} v v^\ddagger \mathcal{M}^\ddagger Z^{-1}f
 &1
}\right) 
	\nonumber\\ 
&&{} \times 
\left(\matrix{
\mathcal{M} v c
 +\frac{1}{2}\mathcal{M}\Delta^\ddagger{}^\alpha
   K_\alpha{}^\beta
   (\varepsilon C)_\beta{}^\gamma \mathcal{M} v \gamma_\gamma
 \cr
 \mathcal{M} v t
 -\frac{1}{2}\mathcal{M}\Delta^\ddagger{}^\alpha
   K_\alpha{}^\beta \mathcal{M} v \gamma_\beta.
}\right)
. 
\label{eq.rs}
\end{eqnarray}

Next, we consider the WZ gauge fixing condition (\ref{eq:C:Necessary}), 
$
 \hat{u}^\ddagger * \partial_\alpha\hat{u}=0
$.
This equation reads
\begin{eqnarray} 
&& 
u^{(0)}{}^\ddagger u^{(1)}{}^\alpha 
= 
	C^{\alpha\beta} u^{(1)}{}^\ddagger_\beta u^{(2)} 
	\label{eq.pgf1} 
	, \\
&& 
u^{(0)}{}^\ddagger u^{(2)} 
= 
	\frac14 u^{(1)}{}^\ddagger{}^\gamma u^{(1)}_\gamma 
	\label{eq.pgf2} 
	, \\ 
&& 
C_{\alpha\beta} 
	u^{(2)}{}^\ddagger u^{(2)} 
= 
	- \frac12 u^{(1)}{}^\ddagger_{(\alpha} u^{(1)}_{\beta)} 
	\label{eq.pgf2'} 
	, \\ 
&& 
u^{(1)}{}^\ddagger_\alpha u^{(2)} 
= 
	 u^{(2)}{}^\ddagger u^{(1)}_\alpha  
	\label{eq.pgf3} 
. 
\end{eqnarray} 
{}From eqs.\ (\ref{eq.c.v2}), (\ref{eq.r_gen}), (\ref{eq.pgf1}) and
(\ref{eq.pgf3})
we find 
\begin{equation}
 \gamma^\alpha
  =\frac{1}{2}
   ({\bf 1}_2 \otimes c^\ddagger
       +2{\cal C} \otimes t^\ddagger)^{-1}{}_\beta{}^\alpha
   u^{(1)\ddagger}{}^{\gamma'}\mathcal{M}^\ddagger{}
   ((\varepsilon C)_{\gamma'}{}^\sigma r_\sigma{}^\beta
      -r_{\gamma'}{}^\sigma (\varepsilon C)_\sigma{}^\beta).
\end{equation}
This equation tells us that $\gamma^\alpha$ vanishes 
if $r_\alpha{}^\beta$
has the form $\delta_\alpha{}^\beta r  + (\varepsilon C)_\beta{}^\alpha s$ 
as in (\ref{eq.v1rs}).
Eq.\ (\ref{eq.v1rs}) with $\gamma^\alpha=0$ 
solves the equations
(\ref{eq.pgf1}) and (\ref{eq.pgf3}).

Then the remaining zero mode equation for $u^{(1)}_\alpha$ gives 
the following two independent equations which correspond to
(\ref{eq.pgf2}) and (\ref{eq.pgf2'}): 
\begin{eqnarray}
c^\ddagger t+
    \left(\matrix{
     r^\ddagger&s^\ddagger
    }\right)
    \left(\matrix{
     \frac{1}{2}f^{-1}Z
     &
     -\frac{1}{4}\det C\mathcal{M} v v^\ddagger \mathcal{M}^\ddagger
     \cr
     -\frac{1}{4}\det C\mathcal{M} v v^\ddagger \mathcal{M}^\ddagger
     &
     -\frac{1}{2}\det Cf^{-1}Z
    }\right)
    \left(\matrix{
     r\cr
     s
    }\right)
    &=&0
, 
\label{eq.pgfcd}\\
      t^\ddagger t-
    \left(\matrix{
     r^\ddagger{}&s^\ddagger
    }\right)
    \left(\matrix{
     -\frac{1}{4}\mathcal{M} v v^\ddagger \mathcal{M}^\ddagger
     &
     -\frac{1}{2}f^{-1}Z
     \cr
     -\frac{1}{2}f^{-1}Z
     &
     \frac{1}{4}\det C\mathcal{M} v v^\ddagger \mathcal{M}^\ddagger
    }\right)
    \left(\matrix{
     r\cr
     s
    }\right)
    &=&0,
\label{eq.pgfcc}
\end{eqnarray}
where $r$ and $s$ are written in terms of $c$ and $t$ as
(\ref{eq.rs}) with $\gamma^\alpha=0$.
We first consider eq.\ (\ref{eq.pgfcd}).
Substituting eq.\ (\ref{eq.rs}) into (\ref{eq.pgfcd}),
we obtain
\begin{eqnarray}
&c^\ddagger t
-\frac{1}{2}
 \left(\matrix{
  c^\ddagger& t^\ddagger
 }\right)
 v^\ddagger \mathcal{M}^\ddagger Z^{-1}f
 \left(\matrix{
  1
  &
  \frac{1}{2}\det C\mathcal{M} v v^\ddagger \mathcal{M}^\ddagger Z^{-1}f
  \cr
  \frac{1}{2}\det C\mathcal{M} v v^\ddagger \mathcal{M}^\ddagger Z^{-1}f
  &
  -\det C
 }\right)\nonumber\\
 &\hspace{20mm}
   \times(1+\frac{1}{4}\det C(\mathcal{M} v v^\ddagger \mathcal{M}^\ddagger Z^{-1}f)^2)^{-1}
     \mathcal{M} v
 \left(\matrix{
  c\cr
  t
 }\right)
 =0
. 
\label{eq.pgfcd_cd}
\end{eqnarray}
To simplify this equation, it is useful to rewrite $t$ as 
\begin{equation}
 t=\frac{1}{2}v^\ddagger \mathcal{M}^\ddagger Z^{-1}f\mathcal{M} v c+t'
\label{eq.dsplit}.
\end{equation}
Using eq.\ (\ref{eq.dsplit}), 
we find that eq.\ (\ref{eq.pgfcd_cd}) becomes
\begin{equation}
 \left(c^\ddagger
  +\frac{1}{2}\det C t^{\prime\ddagger} v^\ddagger \mathcal{M}^\ddagger Z^{-1}f
     (1+\frac{1}{4}\det C(\mathcal{M} v v^\ddagger \mathcal{M}^\ddagger Z^{-1}f)^2)^{-1}
     \mathcal{M} v
     \right)t'=0.
\end{equation}
Note that $c$ cannot be proportional to $\det C$ 
when we consider perturbative solutions, 
because $c=1$ in the undeformed case. 
Thus the bracket in front of $t'$ cannot be zero,
and $t'$ must vanish. 
As a result, we obtain
\begin{equation}
 t=\frac{1}{2}v^\ddagger \mathcal{M}^\ddagger Z^{-1}f\mathcal{M} v c.
\label{eq:C:GenSol:t}
\end{equation}
We can also verify that this satisfies eq.\ (\ref{eq.pgfcc}). 
Now $r$ and $s$ can be obtained by substituting (\ref{eq:C:GenSol:t}) 
into (\ref{eq.rs}): 
\begin{equation}
 r= Z^{-1}f\mathcal{M} v c 
, \quad 
 s=0.
\end{equation} 

So far we have solved the zero mode equation (\ref{eq:app:C:ZeroMode:Comp}) 
and imposed the gauge fixing conditions (\ref{eq.pgf1})--(\ref{eq.pgf3}).
The remaining freedom is $c$.
This is fixed by the normalization condition for the zero mode $\hat{u}$, as
we will see below.
The normalization condition is eq.\ (\ref{eq:C:Normalization}): 
\begin{eqnarray} 
{\bf 1}_n 
&=& 
	u^{(0)}{}^\ddagger u^{(0)} 
	- \frac12 C^{\alpha\beta} u^{(1)}{}^\ddagger_\alpha u^{(1)}_\beta 
	- \det C u^{(2)}{}^\ddagger u^{(2)} 
	\nonumber\\ 
&&{} 
	+ \theta^\alpha  
		\left( 
		u^{(1)}{}^\ddagger_\alpha u^{(0)} 
		+ u^{(0)}{}^\ddagger u^{(1)}_\alpha 
		- \varepsilon_{\alpha\beta} C^{\beta\gamma} 
			( u^{(1)}{}^\ddagger_\gamma u^{(2)} 
			- u^{(2)}{}^\ddagger u^{(1)}_\gamma ) 
		\right) 
	\nonumber\\ 
&&{} 
	+ \theta\theta 
		\left( 
		u^{(2)}{}^\ddagger u^{(0)} 
		+ u^{(0)}{}^\ddagger u^{(2)} 
		- \frac12 u^{(1)}{}^\ddagger{}^\gamma 
			u^{(1)}_\gamma 
		\right) 
. 
\label{eq:C:Normalization:Comp}
\end{eqnarray} 
The only non-trivial condition comes from the lowest component, 
because the higher components in the r.h.s.\ vanish automatically if 
(\ref{eq.pgf1}) and (\ref{eq.pgf2}) are satisfied. 
It leads us to 
\begin{equation}
 c=N^{-\frac{1}{2}}
  =( 1
    +\frac{1}{4}\det C
	(v^\ddagger \mathcal{M}^\ddagger Z^{-1}f\mathcal{M} v)^2)^{-\frac{1}{2}}.
\end{equation}

After some algebra, we obtain the zero mode $\hat{u}$ which satisfies
simultaneously the WZ gauge conditions and the normalization condition:
\begin{eqnarray}
 \hat{u}&=&( 1
         +\Delta^\ddagger{}^\alpha f\mathcal{M}\theta
         +\frac{1}{2}\mathcal{M}^\ddagger f\mathcal{M}\theta\theta
         )u^{(0)}
 \label{eq.c.vhat}
         \\
 u^{(0)}&=&( 1
       -\frac{1}{2}\Delta^\ddagger{}^\alpha K_\alpha{}^\beta
         (\varepsilon C)_\beta{}^\gamma\Delta_\gamma\mathcal{M}^\ddagger
         Z^{-1}f\mathcal{M}
         )v N^{-\frac{1}{2}}U
 \label{eq.c.v0usingv}
         \\
 N&=& 1
    +\frac{1}{4}\det C(v^\ddagger \mathcal{M}^\ddagger
    Z^{-1}f\mathcal{M} v)^2
    \label{eq.c.normalizationNcal}
\end{eqnarray}
where $U$ is a unitary matrix such that $U^\dagger = U^{-1}$, 
and $K, Z, N$ are given by 
(\ref{eq:C:K}), (\ref{eq:C:Z}), (\ref{eq:C:N}), respectively.

\section{${\rm U}(2)$ one instanton}\setcounter{equation}{0}
\label{app:U(2)OneInstanton} 

In this appendix,
we construct deformed super instantons 
in the case of ${\rm U}(2)$ gauge group and instanton number $k=1$ 
with the use of the deformed super ADHM construction 
in section \ref{sect:DSADHM}. 
We will find our solutions are consistent with the results in 
refs.\ \cite{Im,GrRiRo,BrFeLuRe,BiFrPeLe}.

Let us begin with solving the deformed ADHM
constraints (\ref{eq.c.badhm}) (\ref{eq.c.fadhm}) 
and express the constrained ADHM data $a_\alpha$ and ${\cal M}$ 
in terms of unconstrained free parameters.
In the ${\rm U}(2)\ k=1$ case, 
the deformed ADHM constraints (\ref{eq.c.badhm}) become
\begin{eqnarray}
 \sigma_\beta^{i\
  \alpha}\omega_{\alpha\dot{\alpha}}\omega^{\ddagger\dot{\alpha}\beta}
                  +\frac{i}{2}C^i\mathcal{M}\mathcal{M}^\ddagger=0\label{eq.c.bconstk1}
	, \\
 \omega_{\alpha\dot{\alpha}}\mu^{\ddagger\dot{\alpha}}
   +\varepsilon_{\alpha\beta} \mu_{\dot{\alpha}}\omega^{\ddagger\dot{\alpha}\beta}=0
. 
\label{eq.c.fconstk1}
\end{eqnarray}
Here we have defined 
$
C^i\equiv
     \frac{1}{2}\eta^i_{\mu\nu}C^{\mu\nu}
$ 
($(C^i)^\ddagger = - C^i$) 
where $\eta^i_{\mu\nu}$ is the 't~Hooft eta symbol (see appendix \ref{app:Notation}). 
Note that in this section we denote 
the ${\rm U}(2)$ gauge index $u$ in (\ref{def.app.moduli}) 
as a dotted spinor index ($\dot{\alpha}$, $\dot{\beta}$, etc.). 

These constraints are solved by
\begin{equation}
 \mu^{\ddagger\dot{\alpha}} 
  = \varepsilon^{\dot{\alpha}\dot{\beta}} \rho^{-1} \zeta^\alpha 
	\omega_{\alpha\dot{\beta}} 
	, \quad 
 \mathcal{M}^{\prime\ddagger}{}^{\dot{\alpha}} = \bar{\xi}^{\dot{\alpha}}
	, \quad 
 \omega_{\alpha\dot{\alpha}}
  = \omega_\mu i \sigma^\mu_{\alpha\dot{\alpha}} 
, 
\label{eq:C:SolADHMConstr} 
\end{equation}
where 
\begin{equation} 
 \omega_\mu \equiv \left( \matrix{ \rho, & 
	-\frac{1}{8}\rho^{-1}C^i(\bar{\xi}\bar{\xi}-\zeta\zeta) 
	}\right)
, 
\end{equation} 
$\rho^\ddagger = \rho$, $( \zeta_\alpha )^\ddagger = \zeta^\alpha$, 
$( \bar{\xi}_{\dot{\alpha}} )^\ddagger = \bar{\xi}^{\dot{\alpha}}$, 
and the remaining parameter $a'_\mu$ can be arbitrary.  
The parameters $\rho$, $\bar{\xi}_{\dot{\alpha}}$ and $\zeta_\alpha$ 
are unconstrained free parameters 
and correspond to 
the scale, supersymmetry and superconformal moduli, respectively
\footnote{Our superconformal moduli parameter $\zeta$ is
different from the one in refs.
\cite{GrRiRo,BrFeLuRe} 
which corresponds to $\rho^{-1} \zeta$ in our convention. 
}.

Up to now, we have solved the deformed ADHM constraints 
and found that the operator $\hat{\Delta}_\alpha$ is written 
in terms of the unconstrained moduli parameters $\rho$, $\bar{\xi}$ and $\zeta$.
In order to construct the connection $\phi$ of the deformed instanton, 
our next task is to determine 
the normalized zero mode $\hat{u}$ of $\hat{\Delta}_\alpha$, 
or $u^{(0)}$ appearing in eq.\ (\ref{eq.c.phivv0}).

As we have mentioned in the previous section, 
the operator $\Delta_\alpha$ 
(the fermionic moduli ${\cal M}$) 
can be split into 
the $C$-independent part $\Delta_0{}_\alpha$ ($\mathcal{M}_0{}$) 
and the residual $C$-dependent part $\delta \Delta_\alpha$ ($\delta \mathcal{M}$) 
(see eq.\ (\ref{eq:C:Split})), 
and the normalized zero mode $v$ of $\Delta_\alpha$ is given by these quantities 
as in eq.\ (\ref{eq.c.vprime}). 
Because the deformed ADHM constraints are solved by (\ref{eq:C:SolADHMConstr}), 
they are now explicitly written with the use of 
\begin{equation} 
\rho_\mu
\equiv 
	\left(\matrix{
	\rho, & {\bf 0} 
	}\right) 
	, \quad 
\delta\rho_{\mu}
\equiv 
	\left(\matrix{
	0, & -\frac{1}{8}\rho^{-1}C^i(\bar{\xi}\bar{\xi}-\zeta\zeta)
	}\right) 
\end{equation} 
as
\begin{eqnarray}
&& 
 \Delta_0{}_\alpha
  =\left(\matrix{
     \rho_{\alpha\dot{\alpha}}
     &
     y_{\alpha\dot{\beta}}
    }\right)
	, \quad 
 \delta\Delta_\alpha
  =\left(\matrix{
     \delta\rho_{\alpha\dot{\alpha}}
     &
     {\bf 0}
    }\right)
    \label{eq.deltaD} 
	, \\ 
&& 
\mathcal{M}_0{}
  = \left(\matrix{
     \rho^{-1}\zeta^\alpha\rho_{\alpha\dot{\alpha}}
     &
     \bar{\xi}_{\dot{\alpha}}
    }\right)
 \label{fmoduli_u2k1}
	, \quad 
 \delta\mathcal{M}
  = -\rho^{-1}\zeta^\alpha\delta\Delta_\alpha 
, 
\end{eqnarray}
where 
$
\rho_{\alpha\dot{\alpha}}
     \equiv \rho_\mu i\sigma^\mu_{\alpha\dot{\alpha}},\
\delta\rho_{\alpha\dot{\alpha}}
     \equiv \delta\rho_{\mu} i\sigma^\mu_{\alpha\dot{\alpha}}
$. 
Note that here we have absorbed the translation moduli $a'_\mu{}_{[1]\times[1]}$ 
into $y_\mu$.
Then the zero mode $\hat{u}$ can be written 
in terms of $\Delta_0{}_\alpha$, $\delta\Delta_\alpha$, 
${\cal M}_0$, $\delta{\cal M}$ 
and the normalized zero mode $v_0$ of $\Delta_0{}_\alpha$. 
Here we choose 
the ${\rm U}(2)$ $k=1$ instanton in the non-singular gauge (the BPST instanton) 
as $v_0{}$
\footnote{
Instead of this, if we choose the 't~Hooft instanton as $v_0$, 
$$ 
 v_0{}=\left(\matrix{
        |y|(y^2+\rho^2)^{-\frac{1}{2}}\delta^{\dot{\alpha}}_{\ \dot{\beta}}\cr 
        -|y|^{-1}(y^2+\rho^2)^{-\frac{1}{2}}y^{\dot{\alpha}\alpha}\rho_{\alpha\dot{\beta}}
       }\right) 
, 
$$ 
then we can construct the singular deformed super
instantons in ref.\ \cite{BiFrPeLe}.
}: 
\begin{equation}
 v_0{}=\rho^{-1}(y^2+\rho^2)^{-\frac{1}{2}}
	\left(\matrix{
	\rho^{\dot{\alpha}\alpha}y_{\alpha\dot{\beta}}\cr
	-\rho^2\delta^{\dot{\gamma}}_{\ \dot{\beta}}
	}\right)
. 
\label{eq.c.vtil}
\end{equation}
Then from eq.\ (\ref{eq.c.vprime}), 
we obtain $v$ as
\begin{eqnarray}
 v&=& v_0 
    -\rho^{-1}(y^2+\rho^2)^{-\frac{3}{2}}
               \left(\matrix{
                  -\rho^2\delta\rho^{\dot{\alpha}\alpha}y_{\alpha\dot{\beta}}\cr
                   y^{\dot{\gamma}\gamma}\delta\rho_{\gamma\dot{\gamma}'}
                   \rho^{\dot{\gamma}'\gamma'}y_{\gamma'\dot{\beta}}
               }\right)\nonumber\\
  &&\hspace{30mm}   -\frac{1}{16}\det C\zeta\zeta\bar{\xi}\bar{\xi}
                 \rho^{-3}(y^2+\rho^2)^{-\frac{5}{2}}
               \left(\matrix{
                (y^2-2\rho^2)\rho^{\dot{\alpha}\alpha}y_{\alpha\dot{\beta}}\cr
                -3\rho^2y^2\delta^{\dot{\gamma}}_{\ \dot{\beta}}
               }\right)\label{eq.v}.
\end{eqnarray}

Next we express $u^{(0)}$ defined in eq.\ (\ref{eq:C:ZeroMode0}) 
in terms of $v_0$, $\Delta_0{}_\alpha$, $\delta\Delta_\alpha$, 
${\cal M}_0$ and $\delta{\cal M}$.  
Note that in the ${\rm U}(2)$ $k=1$ case, the following equation holds:
\begin{equation}
 (v_0{}^\ddagger \mathcal{M}_0{}^\ddagger \mathcal{M}_0{} v_0{})^2=0.\label{eq.c.vmmv2}
\end{equation}
This equation will turn out to make the almost all quantities vanish in
our calculation.
To verify this equation, we need to use $v_0{}$ in (\ref{eq.c.vtil}). 
Together with the form of $\mathcal{M}_0{}$ in (\ref{fmoduli_u2k1}),
the above equation (\ref{eq.c.vmmv2}) is easily shown.
First, the inverse of $Z$ in (\ref{eq:C:Z}) is found to be 
\begin{equation}
 Z^{-1}
= 
 1
   -\frac{1}{4}\det C f_0 \mathcal{M}_0 \Delta_0^\ddagger{}^\alpha
     f_0 \mathcal{M}_0 \mathcal{M}_0^\ddagger 
	f_0 \Delta_0{}_\alpha \mathcal{M}_0^\ddagger 
	+ \cdots
\end{equation}
by using eq.\ (\ref{eq:C:K}), 
where $\cdots$ denotes the terms depending on 
more than four fermionic moduli. 
Note that in the case of $\textrm{U}(2)$ $k=1$, 
there are only four fermionic moduli parameters, 
so that the ${\cal O}({\cal M}_0{}^5)$ terms vanish automatically. 
Here we have used the equation 
${\cal M} \Delta^\ddagger{}^\alpha f 
(\varepsilon C)_\alpha{}^\beta \Delta_\beta  {\cal M}^\ddagger = 0
$, 
which follows from the fermionic ADHM constraint, 
the symmetric property of $C^{\alpha\beta}$  
and the fact that $\mathcal{M}\Delta^\ddagger{}^\alpha$ and
$\Delta_\alpha\mathcal{M}^\ddagger$ anticommute in the $k=1$ case. 
Then, due to eq.\ (\ref{eq.c.vmmv2}), 
the normalization factor $N$ (\ref{eq:C:N}) becomes 
$N={\bf 1}_2+\mathcal{O}(\mathcal{M}^8)$.
As a result, 
we find an expression of $u^{(0)}$ from eq.\ (\ref{eq:C:ZeroMode0}) as 
\begin{equation}
 u^{(0)}
  =\bigg(
     {\bf 1}_{4} 
    -\frac{1}{2}
      \Delta^\ddagger{}^\alpha f (\varepsilon C)_\alpha{}^{\beta}\Delta_\beta
             \mathcal{M}^\ddagger f \mathcal{M}
 \\
    -\frac{1}{4}\det C 
          \Delta_0^{\ddagger\alpha} f_0\mathcal{M}_0
             \mathcal{M}_0^\ddagger f_0 \Delta_0{}_\alpha
                 \mathcal{M}_0^\ddagger f_0 \mathcal{M}_0
   \bigg)v.
   \label{eq.c.v0u2k1}
\end{equation}
Here we have taken $U={\bf 1}_2$ for simplicity.

Substituting eq.(\ref{eq.c.v0u2k1}) with eq.(\ref{eq.c.vprime}) 
into the eq.(\ref{eq.c.phivv0}),
we obtain the connection  
superfield after a lengthy but straightforward calculation:
\begin{eqnarray}
  (\phi_\mu)^{\dot{\beta}}{}_{\dot{\gamma}}
   &=&-2(y^2+\rho^2)^{-1}\bar{\sigma}_{\mu\nu}{}^{\dot{\beta}}{}_{\dot{\gamma}}
	y_\nu\nonumber
   \\
   &&
    +\frac{1}{4}C_{\mu\nu}\partial_\nu
               \left( K_1\bar{\xi}\bar{\xi}
                     +K_2\zeta\zeta
                     +2\rho K_3
               \right)
    -\frac{1}{4}\det C\rho^{-2}(y^2+\rho^2)^{-2}
         \zeta\zeta\bar{\xi}\bar{\xi}
         \bar{\sigma}_{\mu\nu}{}^{\dot{\beta}}{}_{\dot{\gamma}}y_\nu\nonumber
 \\
 &&
  +\rho^2(y^2+\rho^2)^{-2}
     (\rho^{-1}\zeta^{\gamma}y_{\gamma\dot{\gamma}'}
            +\bar{\xi}_{\dot{\gamma}'})
   (\varepsilon^{\dot{\beta}\dot{\gamma}'}\varepsilon_{\dot{\gamma}\dot{\alpha}}
       +\delta^{\dot{\beta}}{}_{\dot{\alpha}}\delta^{\dot{\gamma}'}{}_{\dot{\gamma}})
   \bar{\sigma}_{\mu}^{\dot{\alpha}\alpha}\theta_\alpha,
\end{eqnarray}
where
\begin{equation} 
 K_1=\frac{y^2}{(y^2+\rho^2)^2}-\frac{2}{y^2+\rho^2},
  \quad
 K_2=\frac{y^2}{(y^2+\rho^2)^2}+\frac{1}{y^2+\rho^2},
  \quad
 K_3=\frac{\zeta\sigma_\sigma\bar{\xi}y^\sigma}{(y^2+\rho^2)^2}.
\end{equation}
{}From this connection superfield, we can read the fermion zero mode and the gauge field
of the deformed super instanton:
\begin{eqnarray}
 (\bar{\lambda}_{\dot{\alpha}})^{\dot{\beta}}{}_{\dot{\gamma}}
  &=&-2i\rho^2(y^2+\rho^2)^{-2}
    (\rho^{-1}\zeta^{\gamma}y_{\gamma\dot{\gamma}'}
            +\bar{\xi}_{\dot{\gamma}'})
   (\varepsilon^{\dot{\beta}\dot{\gamma}'}\varepsilon_{\dot{\gamma}\dot{\alpha}}
       +\delta^{\dot{\beta}}{}_{\dot{\alpha}}\delta^{\dot{\gamma}'}{}_{\dot{\gamma}})
   , \\
 (v_\mu)^{\dot{\beta}}{}_{\dot{\gamma}}
  &=&
   -4i(y^2+\rho^2)^{-1}\bar{\sigma}_{\mu\nu}{}^{\dot{\beta}}{}_{\dot{\gamma}}y_\nu
    +\frac{i}{2}C_{\mu\nu}\partial_\nu
               \left( K_1\bar{\xi}\bar{\xi}
                     +K_2\zeta\zeta
                     +2\rho K_3
               \right) \delta^{\dot{\beta}}{}_{\dot{\gamma}} \nonumber
 \\
&&{} 
    -\frac{i}{2}\det C\rho^{-2}(y^2+\rho^2)^{-2}
         \zeta\zeta\bar{\xi}\bar{\xi}
         \bar{\sigma}_{\mu\nu}{}^{\dot{\beta}}{}_{\dot{\gamma}}y_\nu
. 
 \label{eq.u2k1gaugefield}
\end{eqnarray}
Note that the fermion zero mode coincides with the
one in the undeformed case, which is consistent with the result in refs.\ 
\cite{GrRiRo,BrFeLuRe}. 
On the other hand, the above gauge field does not coincide with the one
in refs.\ \cite{GrRiRo,BrFeLuRe} 
by the term proportional to $\det C$. 
This does not mean that our result contradicts the known results. 
As we have shown in the previous sections, 
our deformed ADHM construction correctly gives the solutions 
to the deformed ASD equations. 
We should note, however, that there is a freedom 
how we parameterize the instanton moduli space. 
In fact, by re-parameterizing the scale parameter in our solution as
\begin{equation}
 \rho \to \rho\left(
                1
               -\frac{1}{16}\det
               C\rho^{-4}\zeta\zeta\bar{\xi}\bar{\xi}
               \right)
, 
\end{equation}
we find that our solution becomes consistent with the one 
obtained in refs.\ \cite{GrRiRo,BrFeLuRe}.

\end{document}